\documentclass[
  aps,prd,
   reprint,
  longbibliography,
    nofootinbib,
  superscriptaddress
]{revtex4-2}

\usepackage[T1]{fontenc}
\usepackage[utf8]{inputenc}
\usepackage[english]{babel}
\usepackage{bm}
\usepackage{graphicx}
\usepackage{xcolor}
\usepackage{amsmath,amssymb}
\usepackage{euscript}
\usepackage{mathrsfs}
\usepackage{soul}
\usepackage{svg}
\usepackage{epstopdf}
\usepackage{hyperref}
\usepackage{slashed}

\newcommand{\dd}{\,d}

\begin{document}

\title{Analytical calculation of the spectrum of nonlinear Compton scattering beyond local approximations}

\author{M.~P. Malakhov}\email{mak.malahov2000@gmail.com}
\affiliation{National Research Nuclear University MEPhI, Moscow, 115409, Russia}
\affiliation{Skolkovo Institute of Science and Technology, 
Skolkovo, 121205, Russia}
\affiliation{
Institute of Applied Physics of the Russian Academy of Sciences, Nizhny Novgorod 603950, Russia}
\author{Th. Benahmed}
\altaffiliation[Current address: ]{Deutsches Elektronen--Synchrotron DESY, Notkestr. 85, 22607 Hamburg, Germany}
\affiliation{ELI Beamlines facility, The Extreme Light Infrastructure ERIC,
Dolni Brezany 252 41, Czech Republic}
\author{E.~G. Gelfer}
\affiliation{ELI Beamlines facility, The Extreme Light Infrastructure ERIC,
Dolni Brezany 252 41, Czech Republic}
\author{A.~M. Fedotov}
\affiliation{National Research Nuclear University MEPhI, Moscow, 115409, Russia}
\affiliation{
Institute of Applied Physics of the Russian Academy of Sciences, Nizhny Novgorod 603950, Russia}
\author{O.~Klimo}
\affiliation{ELI Beamlines facility, The Extreme Light Infrastructure ERIC,
Dolni Brezany 252 41, Czech Republic}
\affiliation{FNSPE, Czech Technical University in Prague, Prague, Czech Republic}
\author{S.~Weber}
\affiliation{ELI Beamlines facility, The Extreme Light Infrastructure ERIC,
Dolni Brezany 252 41, Czech Republic}
\author{S.~G. Rykovanov}
\affiliation{Skolkovo Institute of Science and Technology, 
Skolkovo, 121205, Russia}
\affiliation{
Institute of Applied Physics of the Russian Academy of Sciences, Nizhny Novgorod 603950, Russia}


\begin{abstract}
We derive compact analytical formulae for the spectrum of nonlinear Compton scattering in a finite plane-wave pulse with a smooth temporal envelope. The strong-field QED probability is reduced to finite-pulse phase integrals, which are evaluated asymptotically for multicycle pulses with a broad class of smooth envelopes. We use the uniform approximation to remove the caustic divergences that appear at the nonlinear edges of broadened harmonics. Away from the caustics, it reduces to the standard saddle-point result. The behavior near the linear edge is further improved by an envelope-corrected saddle-point approximation. The approach retains the harmonic substructure in the spectral-angular region carrying the dominant part of the emitted radiation. The locally monochromatic approximation is recovered by averaging the finite-pulse interference. Within their asymptotic domain of applicability, the resulting formulae agree with direct numerical calculations and can be used to evaluate spectra from an electron beam.
\end{abstract}

\maketitle

\section{Introduction}
\label{sec:introduction}

Collisions of high-intensity laser pulses with relativistic electron beams
provide a controlled platform for studying radiation emission in strong
electromagnetic backgrounds \cite{fedotov2023advances, di2012extremely, gonoskov2022charged}. They are also the basis of inverse Compton
sources, where the Doppler upshift converts optical laser photons into
tunable x-ray and $\gamma$-ray radiation
\cite{leemans1996x,albert2010characterization,rykovanov2014quasi,khrennikov2015tunable,graves2014compact,ta2012all,yan2017high}.
Such radiation is attractive for medical imaging
\cite{tashima2022compton}, ultrafast radiography
\cite{tommasini2011development}, studies of nanoscopic
structure \cite{kulpe2020spectroscopic}, and photonuclear applications
\cite{nedorezov2021nuclear}. In strong laser fields the emitted radiation is shaped by the nonlinear motion
of the electron in the wave~\cite{sarachik1970classical,esarey1993nonlinear,popruzhenko2023dynamics}. This nonlinearity modifies the spectral and
angular structure of the radiation \cite{bula1996observation} and, when photon emission becomes
sufficiently strong, can also give rise to radiation-reaction effects \cite{blackburn2020radiation,bulanov2004interaction,gonoskov2022charged,gelfer2018unexpected,gelfer2018theory,gelfer2021radiation,di2010quantum,vranic2016quantum,cole2018experimental,poder2018experimental,los2026observation}. The
high-energy photons produced in the same collisions may also participate in electron--positron pair production in more extreme regimes, including trident and nonlinear Breit–Wheeler channels in pulsed strong-field backgrounds \cite{burke1997positron, bamber1999studies,mironov2021onset,bell2008possibility,fedotov2010limitations,gelfer2015optimized,mercuri2025growth, hu2010complete, mackenroth2018nonlinear, tang2021pulse}.

The physical regime is commonly characterized by two invariant parameters
\cite{nikishov1964quantum1}. The classical nonlinearity parameter is
$a_0=eE_0/(m\omega_L)$, where $-e$ and $m$ are the electron charge and mass,
and $E_0$ and $\omega_L$ are the characteristic laser field amplitude and
frequency. We use units $\hbar=c=1$, and denote scalar products of
four-vectors by $a\cdot b=a^\mu b_\mu$. Classically, $a_0$ measures the
electron quiver momentum in units of $m$. In the quantum description it
measures the strength of the coupling to the coherent background and controls
the importance of multiphoton exchange with the laser wave
\cite{popruzhenko2023dynamics}. For $a_0\ll1$, the process enters the linear Compton regime. For $a_0\gtrsim1$, the interaction becomes nonlinear and higher harmonics appear in the emitted spectrum.

The second invariant parameter is $\chi=e\sqrt{-(F_{\mu\nu}p^\nu)^2}/m^3$, where $F_{\mu\nu}$ is the electromagnetic field tensor and $p^\mu$ is the electron four-momentum \cite{nikishov1964quantum1, ritus1985quantum}. It controls the importance of quantum recoil. The regime $a_0\gtrsim1$ with $\chi\ll1$ is nonlinear but essentially classical and corresponds to nonlinear Thomson scattering. When $\chi$ becomes comparable to unity, the emitted photon can carry away a substantial fraction of the electron energy, and a quantum treatment of the emission process is required. The relation between nonlinear Thomson and nonlinear Compton spectra in shaped pulses, including spin and polarization effects, has been studied through frequency-scaling laws~\cite{krajewska2014frequency}.

The elementary process considered here is nonlinear Compton scattering, i.e., the emission of a single photon by an electron interacting with an intense laser pulse, schematically $e_L\to e_L'+\gamma'$. It is described within strong-field QED using laser-dressed electron states
\cite{furry1951bound, nikishov1964quantum1, ritus1985quantum,landau4}. This treatment is nonperturbative in the
background field and includes the recoil of the emitted photon. The nonlinear interaction with the laser gives rise to characteristic spectral features, including higher harmonics and the intensity-dependent redshift of the scattered radiation. Such effects were observed in the SLAC E144 experiment \cite{bula1996observation,bamber1999studies}, and recent laser--electron experiments have entered regimes where quantum effects become increasingly important \cite{mirzaie2024, los2026observation}. Detailed knowledge of the radiation spectrum is therefore important for interpreting experiments, benchmarking numerical models, and using the emitted photons as a diagnostic of the laser and electron-beam parameters~\cite{har2012peak, blackburn2020model, he2019towards}.

The structure of the spectrum in a finite plane-wave pulse is more involved than in the monochromatic limit. The distinction between ideal monochromatic waves and physical wave packets was emphasized in Ref.~\cite{dawson1970plane}. In an infinite wave the exact periodicity fixes the net four-momentum exchange with the background to discrete harmonic channels \cite{sarachik1970classical, nikishov1964quantum1,ritus1985quantum}. For a finite pulse this periodicity is no longer exact, and the local intensity changes during the interaction. The laser dressing of the electron therefore varies across the pulse, so the intensity-dependent redshift is not fixed by a single field amplitude. As a result, each monochromatic harmonic is broadened into a finite spectral band
\cite{esarey1993nonlinear,hartemann1996classical,krafft2004spectral, brau2004oscillations, boca2009nonlinear, mackenroth2011nonlinear, seipt2011nonlinear, krajewska2012compton, seipt2013nonlinear}. For a fixed frequency and observation direction, radiation can be emitted at different phases of the envelope, typically on its rising and falling sides. The coherent superposition of these contributions produces subpeak structure within each broadened harmonic and makes the spectrum sensitive  to the pulse-dependent harmonic broadening and to the relative phase between emission points. Related finite-pulse, pulse-envelope, and multipulse interference effects in nonlinear Compton and Thomson scattering have been analyzed in Refs.~\cite{king2021interference, wistisen2014interference, ilderton2020toward, krajewska2014global, zhao2025polarization}. The ponderomotive broadening can be reduced by pulse shaping, for example by chirping \cite{ghebregziabher2013spectral,seipt2019optimizing}, by using suitable pulse profiles \cite{timoshenko2025pulse}, or by polarization gating \cite{valialshchikov2021narrow}.

Exact analytical integration of the corresponding oscillatory phase integrals is possible only for special pulse shapes, such as rectangular pulses \cite{esarey1993nonlinear}, hyperbolic-secant pulses \cite{seipt2016analytical}, or some generalized hyperbolic-secant envelopes \cite{malakhov2025calculation}. For a generic smooth envelope one usually evaluates these integrals numerically \cite{Thomas_2010, kharin2016temporal, malakhov2025calculation} or uses local approximations. The locally constant field approximation (LCFA) \cite{ritus1985quantum,di2018implementing,ilderton2019extended,gelfer2022nonlinear}
is computationally efficient and widely used in simulation codes
\cite{gonoskov2015extended,di2019improved}, but its applicability requires the formation length to be much shorter than the field-variation scale and usually assumes $a_0\gg1$, together with the condition $a_0^3/\chi\gg1$ \cite{di2018implementing,dinu2016quantum}. Since the LCFA replaces the laser field by a locally constant crossed field, it resolves neither the harmonic structure of the spectrum nor the finite-pulse substructure within the harmonics~\cite{harvey2015testing}. For laser pulses with a well-defined carrier frequency and many optical cycles, an important alternative is the locally monochromatic approximation (LMA) \cite{heinzl2020locally}. The LMA uses the separation between the fast carrier oscillations and the slowly varying envelope, is not restricted to very large $a_0$, and can describe local harmonic structures, which makes it especially useful in the transition regime $a_0\sim1$ and in numerical simulations \cite{blackburn2021local,nielsen2022high,blackburn2023ptarmigan, larin2025extended}. However, the LMA is local in phase and cycle averaged, so it does not resolve the long-range interference responsible for finite-pulse subpeaks.

\begin{figure}[t]
    \centering
    \includegraphics[width=0.95\columnwidth]{
    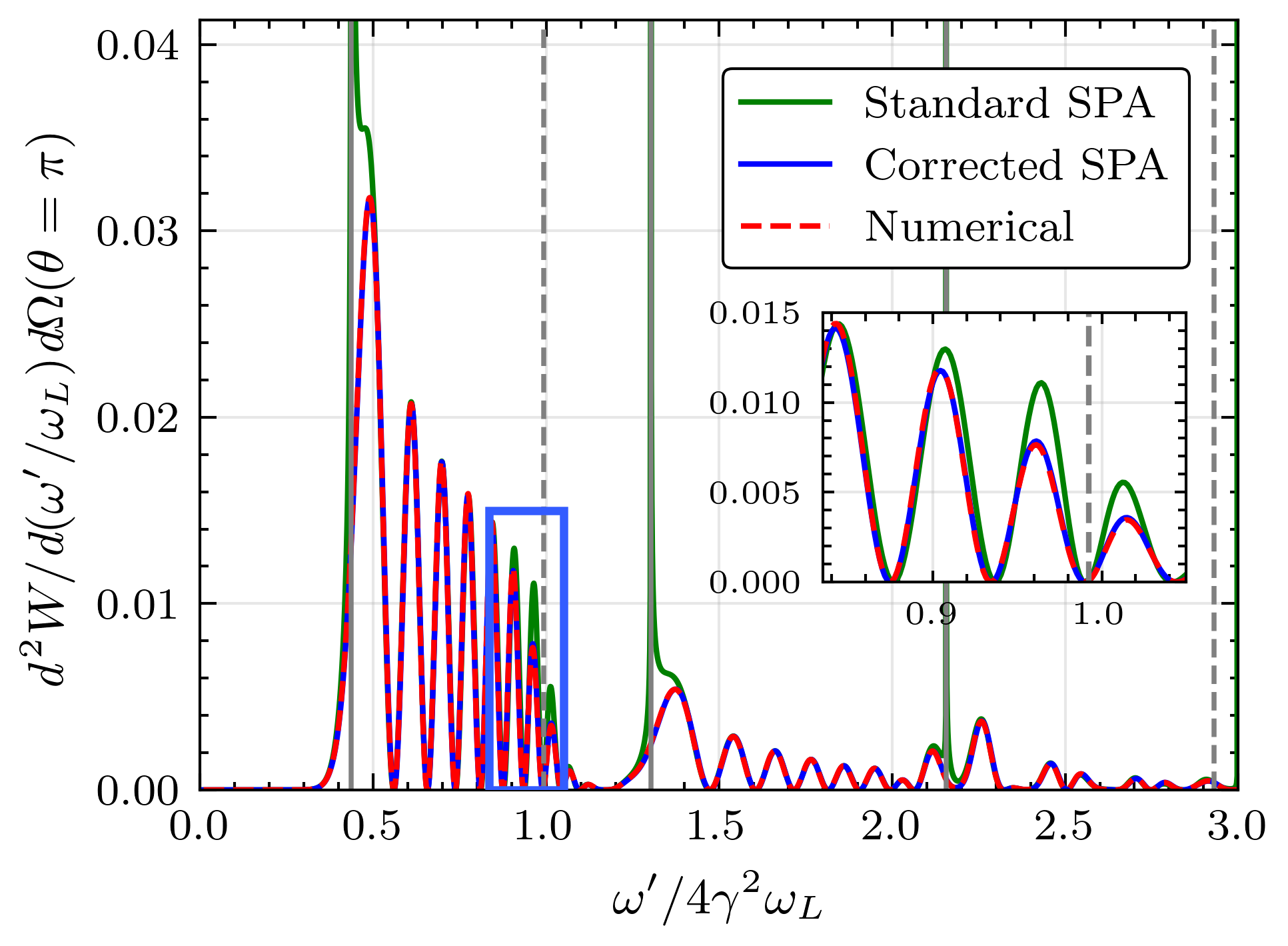}
    \caption{Illustration of the edge problem in the finite-pulse spectrum for a representative set of laser and electron parameters. The green, blue, and red dashed curves show the standard SPA, the corrected SPA, and the numerical calculation, respectively.}
    \label{fig:spec_intro}
\end{figure}

For multicycle pulses, the spectrum can be analyzed using the stationary-phase approximation \cite{seipt2016analytical,kharin2016temporal,narozhnyi1996photon} or, more generally, the saddle-point approximation (SPA). Radiation at a fixed frequency and observation direction is formed mainly at those phases of the pulse where the emission amplitudes interfere constructively. For a smooth pulse profile, two such phases contribute inside a harmonic band and correspond to equal local intensities on the rising and falling sides of the pulse. Their interference gives the subpeak pattern, which is not resolved by the LMA or the LCFA. The standard SPA captures this structure in the interior of the band, but it fails near the harmonic edges, where the saddle-point configuration changes. This behavior is illustrated in Fig.~\ref{fig:spec_intro}. The gray solid vertical lines mark the nonlinear edges. At these points the two saddles coalesce near the pulse maximum, which leads to a divergence of the standard SPA \cite{kharin2018higher}. The gray dashed vertical lines mark the linear edges. Near these edges the standard SPA can overestimate the subpeaks, as shown in the inset of Fig.~\ref{fig:spec_intro}.

In this paper we develop an analytical description of nonlinear Compton scattering in a finite plane-wave pulse that preserves the harmonic structure and regularizes the nonlinear spectral edge. Starting from the SFQED probability, we express the spectral-angular distribution through phase integrals containing the dependence on the pulse shape, polarization, and observation direction~\cite{seipt2011nonlinear,seipt2013asymmetries}. For multicycle pulses these integrals are organized into harmonic-resolved phase integrals. Each harmonic-resolved phase integral is described by a uniform Airy expression, which regularizes the nonlinear edge and reduces to the ordinary saddle-point approximation in the interior of the band. Near the linear edge this description is supplemented by an envelope-corrected SPA to improve the last subpeaks. We also show how the LMA is recovered by averaging the finite-pulse interference in the harmonic-resolved integrals. The resulting formulae are benchmarked against direct numerical calculations for different polarizations, pulse shapes, and observation angles. Finally, we apply the corrected single-electron formulae to incoherent spectra from an electron beam and discuss when the harmonic substructure survives beam averaging.

The remainder of the paper is organized as follows. In Sec.~\ref{sec:scattering_amplitude_probability_energy} we introduce the finite plane-wave background and obtain the SFQED probability in terms of phase integrals. In Sec.~\ref{sec:master} we use the slowly varying envelope approximation to separate the carrier oscillations from the envelope dependence and write the phase integrals in harmonic-resolved form. In Sec.~\ref{sec:asymptotics} we develop the asymptotic description, including the standard saddle-point approximation, the uniform Airy approximation, and the envelope-corrected treatment near the linear edge. In Sec.~\ref{sec:lma} we show how the locally monochromatic approximation is recovered by averaging the finite-pulse interference in the harmonic-resolved integrals. In Sec.~\ref{sec:results} we compare the analytical formulae with numerical calculations and discuss their applicability for single-electron spectra. In Sec.~\ref{sec:beam} we apply the corrected single-electron formulae to incoherent spectra from an electron beam. We summarize our results in Sec.~\ref{sec:conclusion}. Technical details are given in the Appendixes, where we also present the corresponding results for nonlinear Thomson scattering.

\section{NONLINEAR COMPTON SCATTERING IN A FINITE PLANE WAVE}
\label{sec:scattering_amplitude_probability_energy}

To account for the interaction with the strong laser pulse nonperturbatively, we work in the
Furry picture \cite{furry1951bound} and use Volkov states \cite{volkov1935solution} as in- and out-states for the electron. The background
field is treated as a prescribed classical finite-duration plane-wave pulse described by the dimensionless four-potential
\begin{equation}
  a^{\mu}(\varphi)
  = a_{0}g(\varphi)
  \left(
      \varepsilon_{1}^{\mu}\cos\xi\cos\varphi
      +
      \varepsilon_{2}^{\mu}\sin\xi\sin\varphi
  \right),
  \label{eq:laser_potential}
\end{equation}
where $\varphi=k\cdot x$ is the invariant laser phase and $k^\mu$ is the laser four-wavevector.
The dimensional potential is $A^\mu=(m/e)a^\mu$, with $e>0$ denoting the absolute value of the electron charge. The envelope $g(\varphi)$ specifies the finite temporal duration of the pulse.
The limit $g\to1$ corresponds to an infinitely long plane wave, whereas an asymptotically vanishing $g(\varphi)$ describes a pulsed plane wave. We assume that the finite-pulse envelope depends on the phase through the scaled variable
$\varphi/\Delta\phi$, where $\Delta\phi=\omega_L T$ characterizes the dimensionless pulse length, and that it is normalized as $g(0)=1$ and satisfies
$g(\pm\infty)=0$. The polarization four-vectors obey $\varepsilon_{1,2}\cdot k=0$ and
$\varepsilon_i\cdot\varepsilon_j=-\delta_{ij}$. The parameter $\xi$ determines the polarization
state of the laser field: $\xi=0$ or $\pi/2$ corresponds to linear polarization along one of the
transverse axes, while $\xi=\pm\pi/4$ gives circular polarization. With this convention the cycle-averaged intensity in a long pulse is
controlled by the same envelope factor $g^2(\varphi)$ for different polarization states.

The Volkov solutions of the Dirac equation in the background field \cite{ritus1985quantum} are written as
\begin{equation}
    \Psi_{p,r}(x)
    =
    \left(
        1+\frac{e}{2k\cdot p}\slashed{k}\slashed{A}(\varphi)
    \right)
    e^{iS_p(x)}u_{p,r},
    \label{eq:volkov_state}
\end{equation}
where $u_{p,r}$ is a free bispinor with momentum $p$ and spin projection $r$, normalized as
$\bar u_{p,r}u_{p,r'}=2m\delta_{rr'}$, and $\slashed{a}\equiv \gamma_\mu a^\mu$ denotes contraction with Dirac matrices. The Volkov phase is the classical Hamilton--Jacobi
action of an electron in a plane-wave background,
\begin{equation}
    S_p(x)=-p\cdot x -
    \frac{1}{2k\cdot p}
    \int\limits^\varphi
    \left[
        2e\left(p\cdot A(\varphi')\right)
        -
        e^2 A^2(\varphi')
    \right]
    \dd\varphi' .
    \label{eq:volkov_action}
\end{equation}
The first term is the free-particle action, while the integral describes the
phase accumulated due to the interaction with the laser field. The part linear in the laser
potential is associated with the transverse quiver motion of the electron. The part quadratic
in the potential depends on the local intensity of the pulse and represents the longitudinal
drift induced by the wave.

The Volkov states include the interaction of the electron with the laser pulse and are used as
incoming and outgoing electron states in the calculation of the matrix element. The emission
of a photon is then treated as a first-order process on top of this nonperturbative
interaction with the background field. For a photon with four-momentum
$k'^\mu=\omega'n'^\mu$ and polarization vector $\varepsilon_{\lambda'}^{\prime\mu}$, the
corresponding $S$-matrix element reads \cite{landau4}
\begin{equation}
    S_{fi}
    =
    -ie
    \int \dd^4x\,
    \bar{\Psi}_{p',r'}(x)\slashed{\varepsilon}_{\lambda'}^{\prime *}
    \frac{
        e^{ik'\cdot x}\sqrt{4\pi}
    }
    {\sqrt{2\omega'}}
    \Psi_{p,r}(x).
    \label{eq:Sfi_general}
\end{equation}
Since the plane wave depends only on the invariant phase $\varphi=k\cdot x$, the integrations over the remaining light-front coordinates can be performed explicitly. They give the
conservation of the three momentum components $(-,\perp)$ which are not affected by the external
plane wave. The remaining integration is over the laser phase $\varphi$. In this way, the amplitude can be written as a superposition of contributions satisfying the four-momentum conservation law $p+sk=p'+k'$, where $s=k'\cdot p/k\cdot p'$ parametrizes the continuous four-momentum exchange
between the electron and the external plane-wave background. For a prescribed observation direction $n'^\mu$, the
conservation law gives the emitted photon energy as
\begin{equation}
  \omega' = \frac{s\,k\cdot p}{(p+sk)\cdot n'}.
\end{equation}
In the infinite monochromatic limit, the periodicity of the background restricts $s$ to
integer harmonic values. In a finite pulse this periodicity is lost, and $s$ becomes a continuous
spectral variable. The harmonic peaks are then broadened into finite bands, whose structure
is determined by the phase integrals introduced below.

After averaging over the initial electron spin and summing over the final electron spin and
photon polarization, the spectral-angular probability \cite{seipt2011nonlinear, seipt2013asymmetries} is
\begin{multline}
  \frac{\dd^{2} W}{\dd\omega'\dd\Omega}
  =
  \frac{\alpha m^2\omega'}{8\pi^{2}(k\cdot p)(k\cdot p')}
  \Bigg[
      -2|A_0|^2
      \\ +
      \frac{a_0^2}{2}
      \left(
          1+\frac{u^2}{2(1+u)}
      \right)
      \big(
          |A_+|^2
          +
          |A_-|^2
          \\+
          2\cos 2\xi\,\operatorname{Re}\left[A_+A_-^*\right]
          -
          2\operatorname{Re}\left[A_0A_2^*\right]
      \big)
  \Bigg],
  \label{eq:dW_main}
\end{multline}
where $u=(k\cdot k')/(k\cdot p')$, and we use the convention
$\alpha=e^2$, with $\alpha$ being the fine-structure constant.

The corresponding spectral-angular distribution of the emitted energy is obtained by
multiplying the probability by the photon energy,
\begin{equation}
    \frac{\dd^2{\cal E}}{\dd\omega'\dd\Omega}
    =
    \omega'
    \frac{\dd^2 W}{\dd\omega'\dd\Omega}.
    \label{eq:dE_main}
\end{equation}

The phase integrals entering in Eq.~\eqref{eq:dW_main} are defined as~\cite{seipt2013asymmetries}
\begin{multline}
\begin{pmatrix}
  A_0 \\
  A_{\pm} \\
  A_2
\end{pmatrix}
=
\int\limits_{-\infty}^{+\infty}
\dd\varphi\,
e^{
    is\varphi
    +
    i\tilde f(\varphi)
    +
    i\beta\int\limits^\varphi g^2(\varphi')\,\dd\varphi'
}
\\ \times
\begin{pmatrix}
  1 \\
  g(\varphi)e^{\mp i\varphi} \\
  g^2(\varphi)\left(1+\cos 2\xi\cos 2\varphi\right)
\end{pmatrix}.
\label{eq:phase_integrals}
\end{multline}
They contain the complete information about the finite duration of the pulse. The factors
multiplying the exponential originate from the Volkov spin structures, while the exponential
itself is the dynamical phase accumulated by the laser-dressed electron during the emission
process.

The rapidly oscillating carrier-dependent part of the phase is
\begin{multline}
  \tilde f(\varphi)
  = \\
  \alpha_1 \cos{\xi}
  \int\limits^\varphi
  g(\varphi')\cos\varphi'\dd\varphi'
  +
  \alpha_2 \sin{\xi}
  \int\limits^\varphi
  g(\varphi')\sin\varphi'\dd\varphi'
  \\+ 
  \beta\cos 2\xi
  \int\limits^\varphi
  g^2(\varphi')\cos 2\varphi'\dd\varphi'.
  \label{eq:tilde_f}
\end{multline}
The term proportional to $\beta\int g^2\dd\varphi$ in the exponential of Eq.~\eqref{eq:phase_integrals} varies on the envelope scale and determines the ponderomotive shift of the harmonic band.

The coefficients in the phase are
\begin{align}
    \alpha_{1,2}
    &=
    ma_0
    \left(
        \frac{\varepsilon_{1,2}\cdot p'}{k\cdot p'}
        -
        \frac{\varepsilon_{1,2}\cdot p}{k\cdot p}
    \right),
    \label{eq:alpha}
    \\
    \beta
    &=
    \frac{m^2a_0^2}{4}
    \left(
        \frac{1}{k\cdot p'}
        -
        \frac{1}{k\cdot p}
    \right).
    \label{eq:beta}
\end{align}
The parameters $\alpha_{1,2}$ describe the transverse quiver contribution projected onto the two polarization directions. The parameter $\beta$ is proportional to $a_0^2$ and describes the intensity-dependent longitudinal part of the Volkov phase. It is therefore the parameter that controls the nonlinear redshift and the width of a harmonic in a finite pulse. This parameter can be expressed as $\beta=bs$, where 
\begin{equation}
    b = \frac{m^2a_0^2}{4}\frac{(n'\cdot k)}{(k \cdot p)(n' \cdot p)}.
    \label{eq:b}
\end{equation}

The integral $A_0$ has to be treated separately, because its integrand does not contain an explicit envelope factor in the pre-exponential part. As a result, it also includes the field-free contribution of the electron current outside the laser pulse. This contribution is
localized at zero momentum transfer and is irrelevant for the radiation spectrum at $s>0$~\cite{boca2009nonlinear}. For positive $s$, $A_0$ is fixed by gauge invariance of the emission amplitude \cite{ilderton2011trident, dinu2012infrared, seipt2013asymmetries} and can be expressed through the finite pulse-dependent integrals $A_\pm$ and $A_2$:
\begin{equation}
    sA_0
    =
    -\frac{1}{2}\alpha_+A_+
    -
    \frac{1}{2}\alpha_-A_-
    -
    \beta A_2,
    \label{eq:A0_gauge_identity}
\end{equation}
where $\alpha_\pm=\cos\xi\,\alpha_1 \pm i\sin\xi\,\alpha_2 $. In the following, this identity is used to eliminate $A_0$ from the final expressions. This is
convenient because $A_\pm$ and $A_2$ are convergent and can be readily evaluated numerically.

Finally, we specify the geometry used below. The laser pulse is taken to propagate along the
positive $z$ axis with $k^\mu=(\omega_L,0,0,\omega_L)$ and the emitted photon direction $n'^\mu=(1,\sin\theta\cos\psi,\sin\theta\sin\psi,\cos\theta)$. The polar angle $\theta$ is measured from the laser propagation direction. In particular, backscattering of the emitted photon corresponds to $\theta=\pi$.

\section{HARMONIC EXPANSION}
\label{sec:master}

A direct evaluation of the phase integrals $A_{\pm}$ and $A_2$ is complicated
by the rapidly oscillating exponential containing the phase $\tilde f(\varphi)$.
For a multicycle pulse, $\Delta\phi\gg1$, the carrier oscillations occur on
the scale of one laser period, whereas the envelope changes on the much longer
scale $\Delta\phi$. This separation of scales allows one to use the slowly
varying envelope approximation (SVEA)
\cite{narozhnyi1996photon,seipt2016analytical}. In this approximation the
field is treated as locally monochromatic over one optical cycle, while the
integration over the full pulse is kept. Therefore, the method preserves phase
coherence between different parts of the pulse and can describe the
substructure of broadened harmonics. Applying this approximation to the carrier-dependent phase $\tilde f(\varphi)$ in Eq.~\eqref{eq:tilde_f}, and neglecting derivatives of the envelope up to corrections suppressed by powers of $1/\Delta\phi$, one obtains
\begin{equation}
  \tilde f(\varphi)
  \approx
  \bar\alpha g(\varphi)\sin(\varphi-\varphi_0)
  +
  \bar\beta g^2(\varphi)\sin 2\varphi ,
  \label{eq:tilde_f_svea}
\end{equation}
where $\bar\alpha=|\alpha_+|$, $\varphi_0=\arg(\alpha_{+})$, and
$\bar\beta=\beta\cos(2\xi)/2$. The first term describes the quiver part of the Volkov phase, whereas the second term is present only for non-circular polarization and originates from the oscillating part of the intensity-dependent phase.

At this point the analogy with an infinite plane wave becomes useful. In a monochromatic wave the periodic phase is expanded into harmonics, and the spectrum consists of discrete harmonic lines. In a pulse the phase is not globally periodic, but it is approximately periodic
over a short interval where the envelope can be regarded as fixed. One can therefore introduce a local harmonic expansion over a floating interval. We write
\begin{equation}
  e^{i\tilde f(\varphi)}
  =
  \sum_{\ell=-\infty}^{+\infty}B_\ell(\varphi)e^{-i\ell\varphi},
  \label{eq:carrier_harmonic_expansion}
\end{equation}
where
\begin{equation}
  B_\ell(\varphi)
  =
  \frac{1}{2\pi}
  \int\limits_{\varphi-\pi}^{\varphi+\pi}
  e^{i\ell\varphi'+i\tilde f(\varphi')}\dd\varphi' .
  \label{eq:local_harmonic_expansion}
\end{equation}
The coefficients $B_\ell$ vary slowly through their dependence on the local value of the envelope and can be expressed in terms of generalized two-argument Bessel functions~\cite{seipt2013nonlinear}:
\begin{equation}
B_\ell(\varphi)=(-1)^{\ell}e^{i\ell\varphi_0}
\mathcal{J}_{\ell}\left(\bar\alpha g(\varphi), \bar\beta g^2(\varphi)\right),
\label{eq:B_l_bessel}
\end{equation}
where
\begin{equation}
\mathcal{J}_{\ell}(\varphi)=\sum_{k=-\infty}^{+\infty}
(-1)^{k}e^{-2ik\varphi_0}
J_{\ell-2k}\!\left(\bar\alpha g(\varphi)\right)
J_k\!\left(\bar\beta g^2(\varphi)\right).
\label{eq:J_l_def}
\end{equation}
The integer $\ell$ labels the harmonic channel inherited from the monochromatic limit. Although strict periodicity is lost in a finite pulse, this index still organizes the spectrum into broadened harmonic bands.

Using the local harmonic expansion
\eqref{eq:carrier_harmonic_expansion}--\eqref{eq:J_l_def}, we insert the
carrier decomposition into the phase integrals \eqref{eq:phase_integrals} and
rescale the phase variable as $\phi=\varphi/\Delta\phi$. This gives
\begin{multline}
\begin{pmatrix}
  A_{\pm} \\
  A_2
\end{pmatrix}
=
\Delta\phi
\sum_{\ell=0}^{+\infty}
(-1)^{\ell} e^{i\ell\varphi_0} \\ \times
\begin{pmatrix}
  -e^{\mp i\varphi_0}\mathcal{C}_{1,\ell\mp1}^{(\ell)} \\
  \mathcal{C}_{2,\ell}^{(\ell)}+\frac{\cos{2\xi}}{2}\left[e^{-2 i\varphi_0}\mathcal{C}_{2,\ell-2}^{(\ell)}+e^{2 i\varphi_0}\mathcal{C}_{2,\ell+2}^{(\ell)} \right]
\end{pmatrix},
\label{eq:Acirc_main}
\end{multline}
Here the harmonic-resolved phase integrals, cf.~\cite{seipt2016analytical}, are defined as:
\begin{equation}
    \mathcal{C}^{(\ell)}_{n,r}=\int\limits_{-\infty}^{+\infty}\mathcal{J}_{r}(\phi)g^{n}(\phi)e^{i\Delta\phi F_\ell(\phi)}\dd\phi,
    \label{eq:C_l_n_main}
\end{equation}
where $n=1$ for $A_{\pm}$ and $n=2$ for $A_2$. The function $F_\ell$ is the slow part of the phase after the carrier oscillations have been separated:
\begin{equation}
  F_\ell(\phi)
  =
  (s-\ell)\phi
  +
  \beta
  \int\limits^\phi g^2(\phi')\dd\phi'.
  \label{eq:F_l}
\end{equation}
It contains no rapidly oscillating carrier terms.  The first term describes the detuning from the monochromatic harmonic position, while the second term represents the slowly varying ponderomotive phase.

The simplest case is a head-on collision observed in exact backscattering. In this geometry the transverse phase parameter $\bar\alpha=0$, and the circular-polarization result reduces to $A_+=\Delta\phi\,\mathcal{C}_{1,0}^{(1)}$, $A_-=\Delta\phi\,\mathcal{C}_{1,0}^{(-1)}$, and $A_2=\Delta\phi\,\mathcal{C}_{2,0}^{(0)}$. In the positive-frequency region corresponding to the first harmonic, the relevant contribution is carried by $A_+$. The contribution from $A_-$ is exponentially suppressed, while $A_2$ is associated with the low-frequency part of the broadened zeroth harmonic and therefore does not contribute to the first harmonic. This makes the backscattering geometry especially simple, since the spectral structure is controlled by a minimal set of reduced integrals and the behavior near the spectral edges can be isolated without additional harmonic mixing.

\section{ASYMPTOTIC EVALUATION OF THE HARMONIC-RESOLVED PHASE INTEGRALS}
\label{sec:asymptotics}

\subsection{Standard saddle-point approximation}
\label{subsec:standard_saddle}

The harmonic-resolved phase integrals \eqref{eq:C_l_n_main} generally cannot be evaluated in closed form for an arbitrary pulse envelope.
They can be further expanded in powers of the envelope and reduced to a family of master integrals~\cite{seipt2016analytical}. Closed analytical expressions for these master integrals are available for hyperbolic-secant and certain generalized hyperbolic-secant envelopes \cite{seipt2016analytical,malakhov2025calculation}. Separately, the spectrum can be evaluated analytically for a rectangular pulse \cite{esarey1993nonlinear}. For a generic smooth envelope, the phase integrals must instead be evaluated numerically \cite{Thomas_2010,kharin2016temporal,malakhov2025calculation} or treated asymptotically. In the multicycle regime $\Delta\phi\gg1$, the phase is rapidly varying, and the method of steepest descent provides a natural analytical approximation.

The saddle points $\phi_s$ are found from the stationarity condition
$F_\ell'(\phi_s)=0$, which gives 
\begin{equation}
    g(\phi_s)=\pm\sqrt{\frac{\ell-s}{\beta}} .
    \label{eq:saddle_condition}
\end{equation}
Equation~\eqref{eq:saddle_condition} separates the spectral domain of the $\ell$th harmonic into three regions bounded by the nonlinear edge $s_{\ell}=\ell/(1+b)$ and the linear edge $s=\ell$, where $b$ is defined in Eq.~\eqref{eq:b}. The first region $s<s_{\ell}$ lies below the nonlinear edge, the second one $s_{\ell}<s<\ell$ is the interior of the broadened harmonic band, and the third one $s>\ell$ is the finite-pulse tail beyond the linear edge. In what follows we restrict the analysis to smooth envelopes for which each harmonic band is described by a single relevant pair of saddle points. In the first region this pair is purely imaginary and gives an exponentially suppressed contribution. At the nonlinear edge $s=s_{\ell}$ the two saddles coalesce at the maximum of the pulse. In the second region they become real and correspond to two emission phases on the rising and falling sides of the envelope with the same local intensity. As the linear edge $s=\ell$ is approached, the real saddles move into the weak-field tails. In the third region they leave the real axis and become complex, giving the finite-pulse tail beyond the linear edge. This topology includes the Gaussian- and sech-type envelopes used below. Envelopes with a different topology near the maximum require another canonical reduction. For example, for a fourth-order super-Gaussian profile the nonlinear edge is not described by two coalescing saddles, and the uniform approximation used below should be generalized to the corresponding saddle topology.

Expanding the phase in Eq.~\eqref{eq:F_l} to second order
near each contributing saddle in the harmonic-resolved phase integrals \eqref{eq:C_l_n_main}, and evaluating the prefactor $g^{n}$ at the saddle point, one obtains \cite{fedoryuk1977saddle}:
\begin{equation}
    \mathcal{C}_{n, r}^{(\ell)}
    =
    \sum_{\phi_s}
    \sqrt{\frac{2\pi i}{\Delta\phi\,F_\ell''(\phi_s)}}
    \,
    \mathcal{J}_{r}(\phi_s)g^{n}(\phi_s)
    e^{i\Delta\phi F_\ell(\phi_s)},
    \label{eq:DN_standard_saddle}
\end{equation}
where the sum is taken over the relevant saddles lying on the deformed contour of steepest descent. In the second region the contributing pair is real $\phi_s=\pm\phi_0$ and Eq.~\eqref{eq:DN_standard_saddle} reduces to the interference form \cite{seipt2013nonlinear,kharin2016temporal}:
\begin{equation}
    \mathcal{C}_{n, r}^{(\ell)}
    =
    \sqrt{\frac{8\pi}{\Delta\phi\,|F_\ell''(\phi_0)|}}
    \,
    \mathcal{J}_{r}(\phi_0)g^{n}(\phi_0)
    \cos\left(
        \Delta\phi F_\ell(\phi_0)-\frac{\pi}{4}
    \right).
    \label{eq:C_l_n_cos_standard}
\end{equation}
This formula shows that the substructure inside a broadened harmonic originates from the interference of radiation emitted at two phases of the pulse with the same local intensity. The standard saddle-point approximation fails at the edges of the harmonic band for two different reasons. At the nonlinear edge $s=s_{\ell}$, the two saddle points coalesce and $F_\ell''(\phi_s)=0$, so that Eq.~\eqref{eq:C_l_n_cos_standard} develops a divergence \cite{kharin2018higher}. This signals that the two saddles can no longer be treated as independent Gaussian contributions. Near the linear edge $s \approx \ell$, the saddle points move into the tails of the pulse, where the result becomes sensitive to the asymptotic form of the envelope and to the variation of the prefactor $g^{n}(\phi)$.

\subsection{Uniform approximation near the nonlinear edge}
\label{subsec:uniform_nonlinear_edge}

Near the nonlinear edge the standard expression \eqref{eq:C_l_n_cos_standard} must be replaced by an approximation which treats the contributing pair of saddles as a single object. In the first region this pair lies on the imaginary axis. With the branch prescription used below, the steepest-descent contour passes through the saddle in the lower half-plane, which gives the exponentially suppressed contribution. At the nonlinear edge the two saddles coalesce, and in the second region they split onto the real axis. The contour then passes through both real saddles, whose
coherent sum produces the interference pattern inside the harmonic band.

A local cubic expansion around the coalescence point regularizes the divergence at the edge, but remains accurate only in a narrow neighborhood of this point
\cite{narozhnyi1996photon,seipt2013nonlinear}. We therefore use a uniform reduction to the canonical Airy integral
\cite{chester1957extension,Felsen_Marcuvitz_RadiationScattering_1994,
milovsevic2025application}. This construction gives a unified expression which is finite at the nonlinear edge and matches the ordinary two-saddle result when the saddles are well separated. The mapping, the choice of the Airy contour, and the corresponding branches are given in Appendix~\ref{app:uniform}.

The uniform approximation allows the harmonic-resolved phase integrals \eqref{eq:C_l_n_main} to be
written as
\begin{equation}
    \mathcal{C}_{n, r}^{(\ell)}
    =
    \sqrt{
        \frac{8\pi^2}
             {|\zeta''(\phi_0)|}
    }
    \mathcal{J}_{r}(\phi_0)g^{n}(\phi_0)
    \begin{cases}
        \operatorname{Ai}\!\left(|\zeta_0|\right),
        & s<s_{\ell}, \\[2mm]
        \operatorname{Ai}\!\left(-|\zeta_0|\right),
        & s \geq s_{\ell}.
    \end{cases}
    \label{eq:DN_uniform_airy}
\end{equation}
Here $|\zeta_0|$ denotes the positive Airy parameter associated with the canonical mapping. It is fixed by the phase difference between the two saddles,
\begin{equation}
    |\zeta_0|
    =
    \left[
        \frac{3}{2}\Delta\phi\,|F_\ell(\phi_0)|
    \right]^{2/3}.
    \label{eq:zeta_def}
\end{equation}
The function $\zeta(\phi)$ maps the original phase $F_\ell(\phi)$ to the canonical cubic phase, and $\zeta_0=\zeta(\phi_0)$. For $s_\ell<s<\ell$ and $|\zeta_0|\gg1$, the two saddles are well separated,
so that Eq.~\eqref{eq:DN_uniform_airy} reduces to
Eq.~\eqref{eq:C_l_n_cos_standard}. The factor
$\zeta''(\phi_0)=d^2\zeta/d\phi^2|_{\phi=\phi_0}$ is determined by matching the local expansion of the canonical phase to that of the original phase at the saddle point. With the branch chosen continuously from the corresponding steepest-descent contour, this matching gives
\begin{equation}
    \zeta''(\phi_0)\zeta^{1/2}(\phi_0)=i\Delta\phi F_\ell''(\phi_0).
    \label{eq:zeta_second_derivative}
\end{equation}
The explicit construction of $\zeta(\phi)$ and the branch prescription are given in
Appendix~\ref{app:uniform}.

\subsection{Linear edge}
\label{subsec:linear_edge}

The correction near the linear edge has a different origin from the nonlinear-edge issue discussed above. When $s\to \ell$, the ordinary saddle-point condition \eqref{eq:saddle_condition} gives $g(\phi_s)\to0$. Therefore the relevant phases move away from the central part of the pulse and enter the weak-field tails of the envelope. In this region the finite duration of the pulse becomes essential. The spectrum does not terminate abruptly at $s=\ell$. Instead, a finite pulse still gives a tail beyond the linear edge, because the loss of exact periodicity produces a finite spectral bandwidth.
For $s>\ell$ the saddle points are no longer real and the contribution is controlled by complex saddles in the tail region.

The standard saddle-point formula \eqref{eq:DN_standard_saddle} becomes inaccurate in this domain because it treats $g^{n}(\phi)$ as a slowly varying prefactor. This is justified in the central part of the harmonic, where the formation region is short compared with the envelope scale. Close to the linear edge, however, the saddle point is pushed into the tail, where $g^{n}(\phi)$ is no longer a slowly varying prefactor. If this factor is kept outside the exponential, the saddle position is determined incorrectly. As a result, the standard approximation may reproduce the phase oscillations qualitatively but gives inaccurate subpeak amplitudes near the linear edge and in the finite-pulse tail beyond it.

To correct this behavior we include the envelope power into the exponential before applying the saddle-point method. The harmonic-resolved phase integrals \eqref{eq:C_l_n_main} can then be written as
\begin{equation}
    \mathcal{C}_{n, r}^{(\ell)}
    =
    \int\limits_{-\infty}^{+\infty}
    \mathcal{J}_{r}(\phi)e^{\Delta\phi\, q_{\ell, n}(\phi)}\dd\phi,
    \label{eq:DN_qN_form}
\end{equation}
where
\begin{equation}
    q_{\ell, n}(\phi)
    =
    iF_\ell(\phi)
    +
    \frac{n}{\Delta\phi}\ln g(\phi).
    \label{eq:qN_def}
\end{equation}
The corrected saddle points of the phase \eqref{eq:qN_def} are determined by
$q_{\ell, n}'(\phi_0)=0$, which gives
\begin{equation}
    i\left((s-\ell)+\beta g^2(\phi_0)\right)
    +
    \frac{n}{\Delta\phi}
    \frac{g'(\phi_0)}{g(\phi_0)}
    =
    0.
    \label{eq:qN_corr}
\end{equation}
Equation~\eqref{eq:qN_corr} identifies the terms which determine the corrected saddle near the linear edge. The residual detuning from the linear harmonic position, the ponderomotive phase, and the exponential decrease of the envelope enter on the same footing. In general this is a transcendental equation, and an exact analytical solution for arbitrary $s$ is available only for special envelopes. For example, for $g(\phi)=1/\cosh\phi$ the equation can be reduced to an algebraic one, as shown in Appendix~\ref{app:linear_edge}. For a generic smooth envelope we only need the saddle in the vicinity of the
linear edge. In this region it can be found by an analytic approximation and, optionally corrected by the first iteration of Newton's method. The construction used below is given in
Appendix~\ref{app:linear_edge}.

The corrected saddle contributing from the upper half-plane is denoted by $\phi_0$ and is chosen in the first quadrant of the complex plane. The second contribution to the same real integral comes from the conjugate saddle $-\phi_0^*$. This pair is selected as lying on a steepest-descent contour obtained by deformation of the original integration path and gives the leading contribution near the
linear edge. The leading Gaussian approximation to Eq.~\eqref{eq:DN_qN_form} already captures the dominant exponential suppression and the oscillating phase. Near the linear edge, however, the formation region lies in the tail of the pulse, where both the corrected phase $q_{\ell,n}(\phi)$ and the generalized
Bessel factor $\mathcal{J}_\ell(\phi)$ vary appreciably over the width of the saddle contribution. To obtain the first correction in $1/\Delta\phi$, we therefore expand the phase through fourth order and
$\mathcal{J}_\ell(\phi)$ through second order about the corrected saddle $\phi_0$. The quadratic expansion of the prefactor is required because its derivatives contribute at the same asymptotic order as the terms generated by the third and fourth derivatives of the phase. These corrections improve the subpeak amplitudes and the matching to the endpoint behavior. The resulting linear-edge approximation, derived in Appendix~\ref{app:linear_edge}, is
\begin{multline}
    \mathcal{C}_{n, r}^{(\ell)}
    =
    \sqrt{\frac{8\pi}{\Delta\phi |q_2|}}
    |\Sigma|
    e^{\Delta\phi \operatorname{Re} q_0}
    \\
    \times
    \cos\left(\Delta\phi \operatorname{Im} q_0 +
    \arg\left( \sqrt{-\frac{2}{q_2}} \Sigma \right)
    \right),
    \label{eq:DN_linear_main}
\end{multline}
where
\begin{multline}
    \Sigma
    =
    \mathcal{J}_{r, 0}+
    \frac{1}{\Delta\phi}
    \Bigg(-\frac{\mathcal{J}''_{r, 0}}{2 q_2}+\frac{\mathcal{J}'_{r, 0} q_3}{2 q_2^2} \\ + 
    \mathcal{J}_{r, 0}
    \left[
    \frac{q_4}
         {8 q_2^2}
    -
    \frac{5 q_3^2}
         {24 q_2^3}
    \right]
    \Bigg),
    \label{eq:linear_defs_main}
\end{multline}
$q_i=q_{\ell, n}^{(i)}(\phi_0)$ and $\mathcal{J}_{r, 0}^{(i)}=\mathcal{J}_{r}^{(i)}(\phi_0)$ denote the $i$th derivatives evaluated at the corrected saddle point $\phi_0$. The exponential factor in Eq.~\eqref{eq:DN_linear_main} describes the suppression due to the tail of the pulse. The square-root factor gives the local formation scale around the corrected saddle. The cosine represents the interference of the two conjugate saddle contributions. Away from the linear edge, where the envelope changes slowly over the formation region, the logarithmic term in Eq.~\eqref{eq:qN_def} becomes a small correction and the result approaches the ordinary two-saddle approximation. Close to the linear edge $s=\ell$ and in the finite-pulse tail beyond it, the corrected phase prevents the saddle from being placed at an incorrect position in the far tail and provides the proper endpoint behavior.

\section{CONNECTION WITH THE LMA}
\label{sec:lma}

The LMA \cite{heinzl2020locally,larin2025extended} is obtained by replacing the finite-pulse interference that remains in the probability by a locally averaged description. The local monochromatic weights are retained, whereas the interference encoded in the harmonic-resolved phase integrals $\mathcal{C}_{n, r}^{(\ell)}$ is averaged over neighboring subpeaks. In this sense, the LMA preserves the local harmonic structure but removes the long-range interference between emission phases separated on the scale of the
pulse duration. This reduction requires a separation between the fast carrier oscillations and the slow variation of the envelope. The pulse must therefore contain many cycles $\Delta\phi\gg1$, so that the field can be treated locally as a monochromatic wave with the slowly varying amplitude $a_0g(\phi)$. In addition, the relative-phase interval contributing to the local probability must remain short compared with the envelope scale. Consequently, unlike the LCFA, the LMA resolves broadened harmonic bands, but it does not reproduce the finite-pulse subpeak structure within each band.

To make the connection with the present formulation explicit, we consider the interior of the $\ell$th harmonic band $s_{\ell}<s<\ell$, where the two contributing saddle points are real and the harmonic-resolved phase integrals are described by the two-saddle interference formula \eqref{eq:C_l_n_cos_standard}. Substitution of Eq.~\eqref{eq:Acirc_main} into the probability \eqref{eq:dW_main} generates both diagonal products associated with the same harmonic and off-diagonal products involving different harmonic
channels. In the local monochromatic description, the latter are removed by averaging over the rapid finite-pulse oscillations. For each remaining diagonal contribution, averaging over neighboring subpeaks gives $\langle\cos^2\rangle=1/2$, and hence
\begin{equation}
    \left\langle
        \left|\mathcal{C}_{n, r}^{(\ell)}\right|^2
    \right\rangle
    =
    \frac{4\pi}
         {\Delta\phi |F_\ell''(\phi_0)|}
    \mathcal{J}_{r}^2(\phi_0)
    g^{2n}(\phi_0).
\end{equation}
Using this averaged expression in Eq.~\eqref{eq:dW_main}, the LMA spectrum can be written in the common form
\begin{equation}
  \frac{\dd^{2} W_{\mathrm{LMA}}}{\dd\omega' \dd\Omega}
  =
  \frac{\alpha m^2\omega'\Delta\phi}
       {\pi(k\cdot p)(k\cdot p')}
  \sum_{\ell=1}^{+\infty}
  \frac{\mathcal{D}_{\ell}(\phi_0)}
       {|F_\ell''(\phi_0)|},
  \label{eq:lma_spectrum_main}
\end{equation}
where the local harmonic weight $\mathcal{D}_{\ell}$ depends on the laser
polarization. In particular, for linear polarization one obtains
\begin{multline}
\mathcal{D}_{\ell}^{(\mathrm{LP})}(\phi_0)
=
-\mathcal{J}_\ell^2
+
\frac{a_0^2}{4}
g^2(\phi_0)
\left(
    1+\frac{u^2}{2(1+u)}
\right)
\\ \times
(
\mathcal{J}_{\ell-1}^2
+
\mathcal{J}_{\ell+1}^2
+
2\mathcal{J}_{\ell-1}\mathcal{J}_{\ell+1} 
-
\mathcal{J}_{\ell-2}\mathcal{J}_{\ell} \\
-
\mathcal{J}_{\ell}\mathcal{J}_{\ell+2} 
- 
2\mathcal{J}_\ell^2
),
\label{eq:lma_weight_lp}
\end{multline}
whereas for circular polarization
\begin{multline}
\mathcal{D}_{\ell}^{(\mathrm{CP})}(\phi_0)
=-J_\ell^2+\frac{a_0^2}{4}g^2(\phi_0)
\left(1+\frac{u^2}{2(1+u)}\right) \\ \times
\left(J_{\ell-1}^2+J_{\ell+1}^2-2J_\ell^2\right).
\label{eq:lma_weight_cp}
\end{multline}
Here $\mathcal{J}_j=\mathcal{J}_j\!\left(\phi_0\right)$ and $J_j=J_j\!\left(\bar\alpha g(\phi_0)\right)$. Equations~\eqref{eq:lma_spectrum_main}--\eqref{eq:lma_weight_cp} reproduce the LMA spectrum for linear and circular polarization \cite{heinzl2020locally,larin2025extended}. Since the result contains the local phase-density factor $1/|F_\ell''(\phi_0)|$, it inherits the fold-type caustic at the nonlinear edge \cite{kharin2018higher}. The LMA$^+$ construction was proposed to overcome this divergence and to restore part of the finite spectral width of the harmonic lines lost in the local LMA reduction \cite{larin2025extended}. In this approach the sharp local harmonic condition of the LMA is replaced by a finite-width profile. For a Gaussian pulse this can be implemented by introducing a Gaussian window in the interference variable, which smears the LMA harmonic lines and removes the divergence of the fully differential LMA spectrum at the nonlinear edge. This improvement is nevertheless not a universal edge approximation for an arbitrary pulse shape. In its explicit implementation, the finite-width profile depends on the chosen window and is matched to a Gaussian envelope. Moreover, LMA$^+$ remains a local averaged approximation. It improves the finite bandwidth of the local harmonic lines, but does not restore the coherent finite-pulse interference responsible for the individual subpeaks. In contrast, the corrected SPA used here retains the phase integrals explicitly and regularizes the edge regions through the Airy and envelope-corrected saddle approximations.

\section{RESULTS AND APPLICABILITY}
\label{sec:results}

In this section we illustrate the performance of the asymptotic formulae for several representative configurations. The obtained analytical spectra are compared with the direct numerical evaluation of the finite-pulse phase integrals. Unless stated otherwise, the initial electron Lorentz factor is
$\gamma=1000$ and the laser photon energy is $\omega_L=1~\mathrm{eV}$.

\begin{figure*}[t]
    \centering
    \includegraphics[width=0.95\textwidth]{
    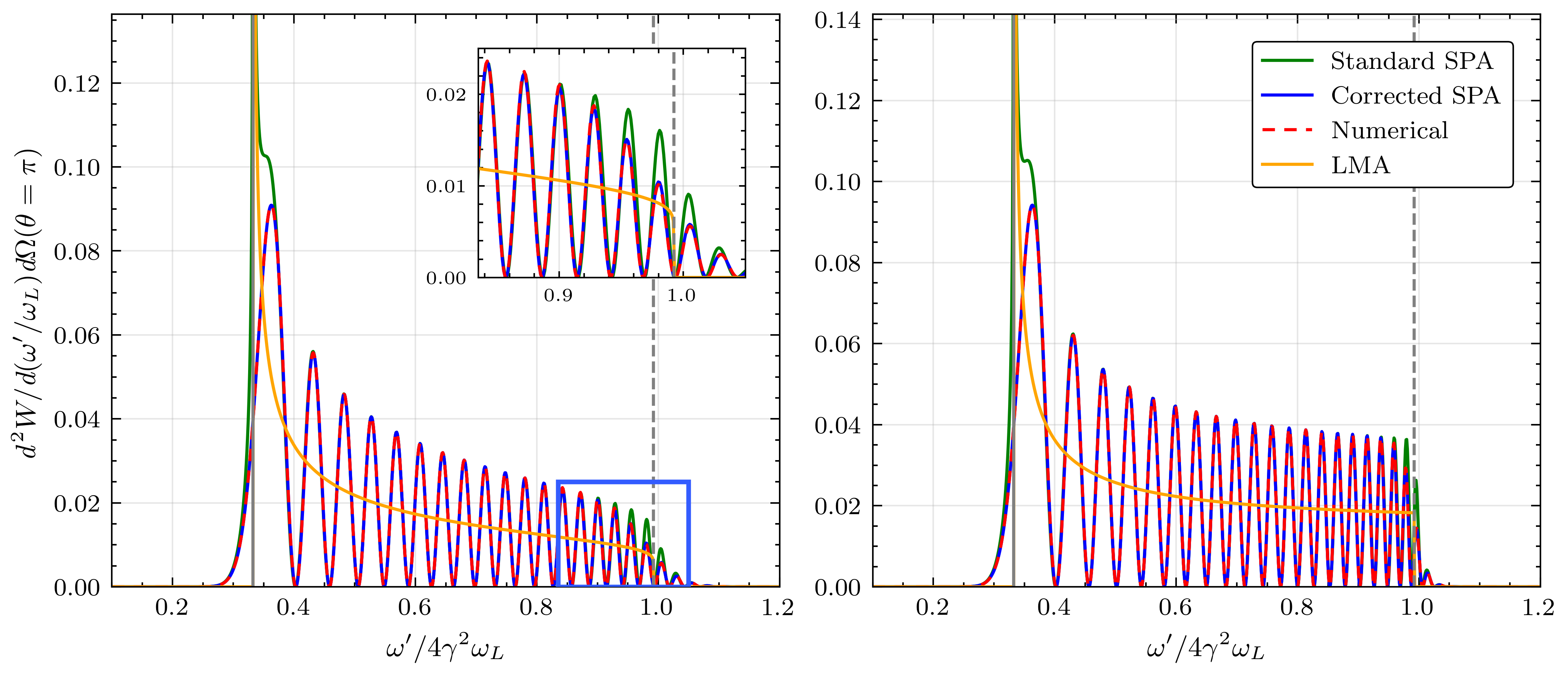}
    \caption{Backscattering spectra for circular polarization with a Gaussian envelope (left panel) and a hyperbolic-secant envelope (right panel). The green curves show the standard SPA, the blue curves the corrected SPA, and the red dashed curves the numerical calculation. The gray solid and dashed vertical lines indicate the nonlinear and linear Compton edges, respectively. The parameters are $a_0=2$, $\Delta\phi=10\pi$, and $\gamma=1000$.}
    \label{fig:backscattering_spectra_circ}
\end{figure*}

We first consider emission in the exact backscattering direction for circular polarization. Figure~\ref{fig:backscattering_spectra_circ} shows the corresponding spectra for Gaussian and hyperbolic-secant envelopes for $a_0=2$ and $\Delta\phi=10\pi$. In this geometry the on-axis spectrum is dominated by the first harmonic. Higher harmonics are emitted predominantly away from the exact backscattering direction.

The nonlinear edge of the $\ell$th harmonic can be written in terms of the
normalized photon frequency $\hat y_\ell$ as
\begin{equation}
    \hat{\omega}'_{\ell}
    =
    \omega_L\hat y_{\ell}\frac{p_-^2}{m^2},
    \qquad
    \hat y_{\ell}
    =
    \frac{\ell}
         {\varkappa_{\ell}
         +a_0^2\sin^2(\theta/2)/2},
    \label{eq:nonlinear_edge_frequency}
\end{equation}
where
\begin{equation}
    \varkappa_{\ell}
    =
    \frac{p_-}{m^2}
    \left(
        p\cdot n'
        +
        2\ell\omega_L\sin^2\frac{\theta}{2}
    \right).
    \label{eq:kappa_edge}
\end{equation}
For an ultrarelativistic particle $p_-^2/m^2 \approx 4\gamma^2$. The
corresponding linear Compton edge is obtained by removing the ponderomotive
shift proportional to $a_0^2$,
\begin{equation}
    \tilde{\omega}'_{\ell}
    =
    \omega_L\tilde y_{\ell}\frac{p_-^2}{m^2},
    \qquad
    \tilde y_{\ell}
    =
    \frac{\ell}{\varkappa_{\ell}} .
    \label{eq:linear_edge_frequency}
\end{equation}
The gray solid vertical lines show the nonlinear edges
\eqref{eq:nonlinear_edge_frequency}, while the gray dashed vertical lines show the linear edges \eqref{eq:linear_edge_frequency}. The nonlinear-edge formula \eqref{eq:nonlinear_edge_frequency} also gives a simple cutoff for the harmonic sum. For a spectrum calculated up to $\omega'_{\max}$, the upper harmonic index can be chosen such that $\hat{\omega}'_{\ell_{\rm cut}}>\omega'_{\max}$.

The first harmonic consists of a dominant peak near the nonlinear edge followed by a sequence of subpeaks. The subpeak pattern is reproduced by the standard SPA expression \eqref{eq:C_l_n_cos_standard} in the interior of the harmonic band, while the nonlinear edge is described by the uniform Airy approximation \eqref{eq:DN_uniform_airy}. The latter also gives an estimate for the position of the first maximum. This estimate is useful because the first maximum fixes the characteristic frequency of the fundamental backscattered peak, which provides the largest contribution to the coherent raidation and coherently enhanced radiation friction for dense electron bunches \cite{gelfer2024coherent, gelfer2025coherently, gelfer2026analytical}. Equating the Airy argument \eqref{eq:zeta_def} to the first maximum with
$|x_1|\approx 1.0188$ \cite{abramowitz1964handbook}, and expanding near the nonlinear edge, one can estimate the position of the first peak of the first
harmonic as $\bar{\omega}_1=\hat{\omega}_1(1+\epsilon)$, where the nonlinear-edge frequency $\hat{\omega}_1$ is defined by Eq.~\eqref{eq:nonlinear_edge_frequency}. The relative shift is
\begin{equation}
    \epsilon
    =
    |x_1|
    \left(
        1-2\frac{\omega_L}{m}\hat y_1\sin^2\frac{\theta}{2}
    \right)^{2/3}
    \left[
        \frac{g''(0)\left(\hat y_1\varkappa_1-1\right)}
             {\Delta\phi^2}
    \right]^{1/3}.
    \label{eq:first_peak_position}
\end{equation}
This shift decreases with increasing pulse duration and is controlled by the curvature of the envelope at its maximum.

The inset in the left panel magnifies the region near the linear edge. In this region the difference between the standard SPA and the corrected SPA is most visible. The standard SPA gives an inaccurate decrease of the subpeaks near the linear edge, whereas the corrected phase $q_{\ell, n}$ in Eq.~\eqref{eq:qN_def} and the corresponding approximation
\eqref{eq:DN_linear_main} improve both the amplitudes and the approach to the
finite-pulse tail.

The corrected SPA curve is obtained by matching the two edge approximations
inside the harmonic band. We use the Airy expression \eqref{eq:DN_uniform_airy} near the nonlinear edge and the envelope-corrected result \eqref{eq:DN_linear_main} near the linear edge, matching them at $s=\ell-\beta/2$ in the calculations below. For sufficiently broad harmonics
the result is insensitive to variations of the matching point.

The LMA curve follows the smooth averaged scale of the spectrum but does not
reproduce the individual subpeaks. It also retains the spurious caustic singularity at the nonlinear edge, as follows from Eq.~\eqref{eq:lma_spectrum_main}. The corrected SPA spectrum remains finite at this edge and follows the numerical result over the harmonic region much more accurately.

\begin{figure*}[t]
    \centering
    \includegraphics[width=0.95\textwidth]{
    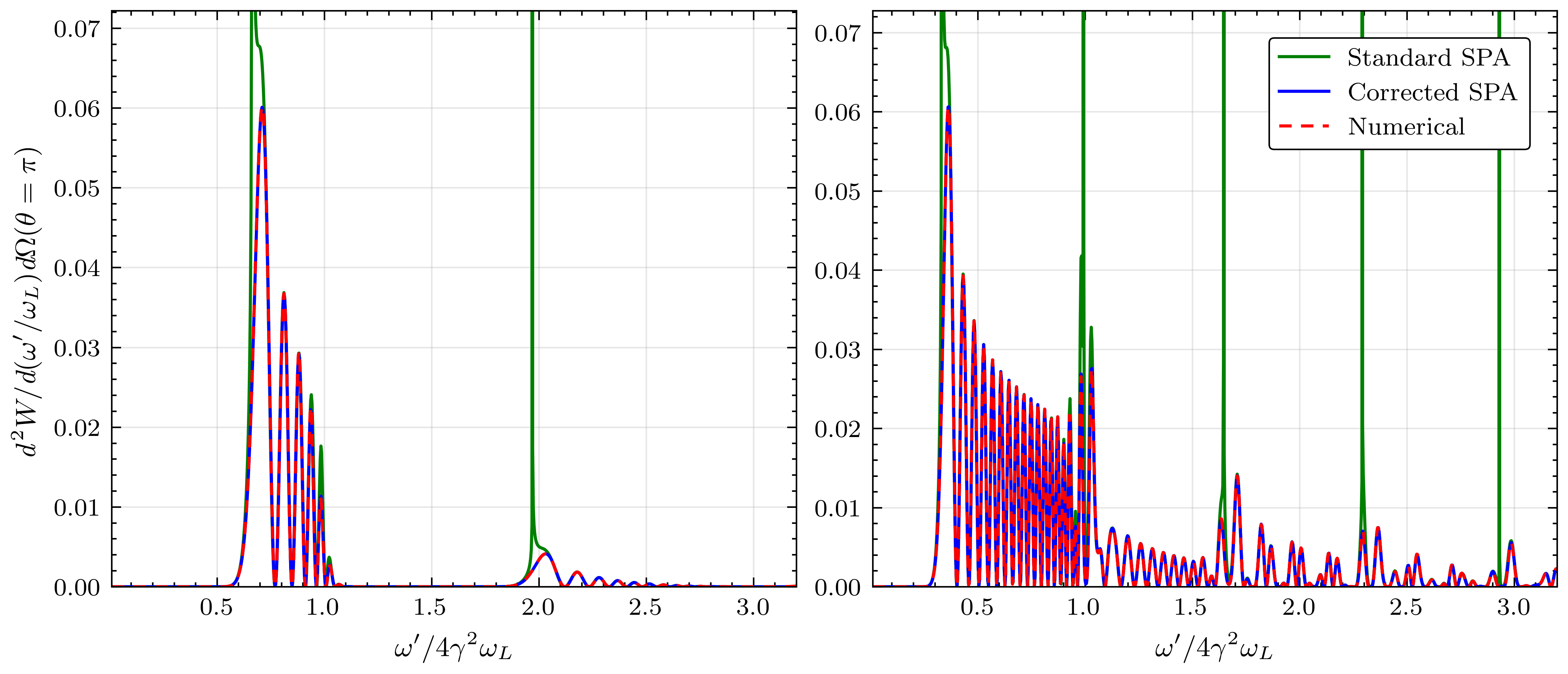}
    \caption{Backscattering spectra for linear polarization with a Gaussian
    envelope. The left and right panels correspond to $a_0=1$ and $a_0=2$,
    respectively. In both cases $\Delta\phi=10\pi$ and $\gamma=1000$. The green, blue, and red dashed curves show the standard SPA, the corrected SPA, and the numerical calculation, respectively.}
    \label{fig:1_spectra_lin}
\end{figure*}

Figure~\ref{fig:1_spectra_lin} shows backscattering spectra for linear
polarization with a Gaussian envelope and $\Delta\phi=10\pi$. In contrast to
circular polarization, higher harmonics are present already in the exact backscattering direction. For $a_0=1$, shown in the left panel, the harmonic bands are well separated and the subpeak structure inside the first harmonic is clearly resolved. When the intensity is increased to $a_0=2$, the ponderomotive broadening becomes stronger and neighboring harmonic bands start to overlap, as shown in the right panel. The standard SPA still reproduces the oscillatory structure inside the harmonics, but shows spurious singular peaks at the nonlinear edge of each harmonic. Since the position of the nonlinear edge depends on the observation angle, these singularities are shifted when one moves away from exact backscattering. Therefore, after angular integration, the same caustic artifacts can contaminate a continuous range of photon frequencies. The corrected SPA removes these singularities and follows the numerical spectrum more accurately, even in the region where the harmonic
bands begin to overlap.

For backscattering one can also estimate how the first harmonic changes with the ellipticity of the laser field. In a head-on collision observed in backscattering, the first harmonic for circular polarization is mainly determined by $A_+^{(\text{CP})}=\Delta\phi\,\mathcal{C}_{1, 0}^{(1)}$, as discussed at the end of Sec.~\ref{sec:master}. For linear and, more generally, elliptic polarization the first harmonic receives the leading contribution from both $A_+$ and $A_-$, whereas $A_2$ contributes to the even harmonic channels, see Eq.~\eqref{eq:Acirc_main}. In the region where the first and second harmonics are weakly overlapping, one may use $A_+ \approx J_{1}(\eta_{\xi})A_+^{(\text{CP})}$ and $A_- \approx J_{0}(\eta_{\xi}) A_+^{(\text{CP})}$, where
\begin{equation}
    \eta_{\xi}
    =
    \frac{\cos(2\xi)}{2}
    \left(
        1-
        \frac{\omega'}{\omega_L}
        \frac{p_+}{p_- - 2\omega'}
    \right).
\end{equation}
Substitution of these expressions into the probability formula \eqref{eq:dW_main} gives the following relation between the first-harmonic spectrum for elliptic polarization and the corresponding circular-polarization spectrum,
\begin{equation}
    \left.
    \frac{\dd^2 W^{(1)}}{\dd\omega'\dd\Omega}
    \right|_{\xi}
    \approx
    \mathcal{R}_\xi
    \left.
    \frac{\dd^2 W^{(1)}}{\dd\omega'\dd\Omega}
    \right|_{\xi=\pi/4},
\end{equation}
where
\begin{equation}
    \mathcal{R}_\xi
    =J_0^2(\eta_{\xi})+J_1^2(\eta_{\xi}) -2\cos(2\xi)J_0(\eta_{\xi})J_1(\eta_{\xi}).
\label{eq:first_harmonic_ellipticity_factor}
\end{equation}
For large $a_0$, neighboring harmonics overlap more strongly for noncircular
polarization. In that case $\mathcal R_\xi$ should be understood as an
estimate of the ratio between the spectrum near the first peak of the first
harmonic and the corresponding circular-polarization spectrum. In the limit
$a_0\gg1$ and in the vicinity of this peak, where $\lambda_1\ll1$, one may
roughly estimate the Bessel functions as
$J_0(\eta_\xi)\approx1$ and $J_1(\eta_\xi)\approx\eta_\xi/2$. Then
Eq.~\eqref{eq:first_harmonic_ellipticity_factor} gives the simple estimate $\mathcal R_\xi\approx 1-\cos^2(2\xi)/2$. Introducing the ellipticity parameter $\varepsilon=\tan\xi$ used in
Ref.~\cite{gelfer2026analytical}, this can be written as
\begin{equation}
    \mathcal R_\xi
    \approx
    1-\frac{1}{2}
    \left(
        \frac{1-\varepsilon^2}{1+\varepsilon^2}
    \right)^2 .
\end{equation}

\begin{figure*}[t]
    \centering
    \includegraphics[width=0.95\textwidth]{
    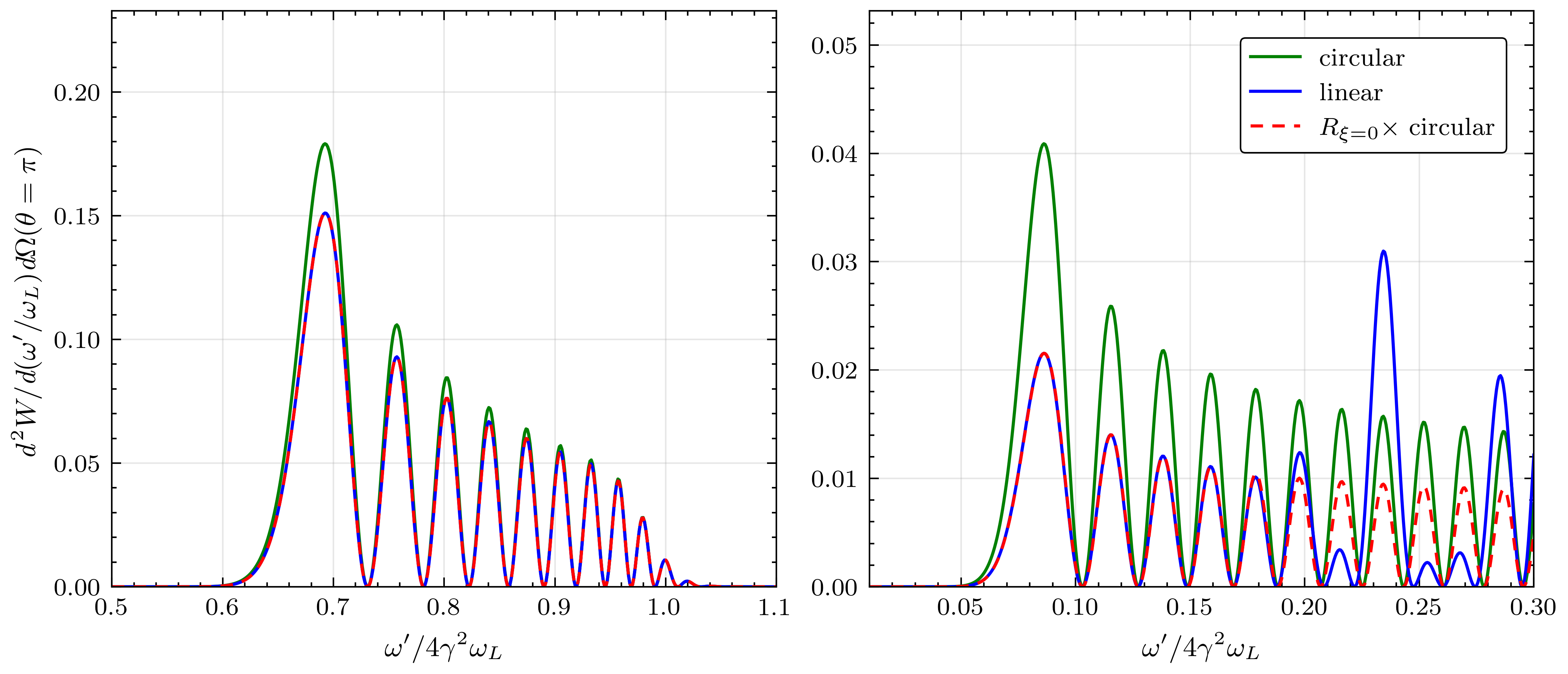}
    \caption{Backscattering spectra for a Gaussian envelope. The green and blue curves correspond to circular and linear polarization, respectively. The red dashed curve shows the circular-polarization spectrum rescaled by $R_{\xi=0}$. The left panel corresponds to $a_0=1$ and $\Delta\phi=20\pi$, while the right panel corresponds to $a_0=5$ and $\Delta\phi=5\pi$. In both cases $\gamma=1000$.}
    \label{fig:spectra_diff_xi}
\end{figure*}

Figure~\ref{fig:spectra_diff_xi} illustrates the scaling
\eqref{eq:first_harmonic_ellipticity_factor} for exact backscattering. The
left panel shows the weakly nonlinear case, $a_0=1$ and $\Delta\phi=20\pi$, where the first harmonic is well separated from higher
ones. In this regime the circular-polarization spectrum rescaled by
$R_{\xi=0}$ reproduces the linear-polarization spectrum over the whole first-harmonic region. This confirms that the polarization dependence of the first harmonic is mainly controlled by the local factor
\eqref{eq:first_harmonic_ellipticity_factor}. The right panel shows a stronger-nonlinear and shorter-pulse case, $a_0=5$ and $\Delta\phi=5\pi$. Here the harmonic bands are broader, and the linear-polarization spectrum receives an additional contribution from the next
odd harmonic. As a result, the scaling by $R_{\xi=0}$ remains reliable only near the first peak of the first harmonic.

\begin{figure}[t]
    \centering
    \includegraphics[width=0.95\columnwidth]{
    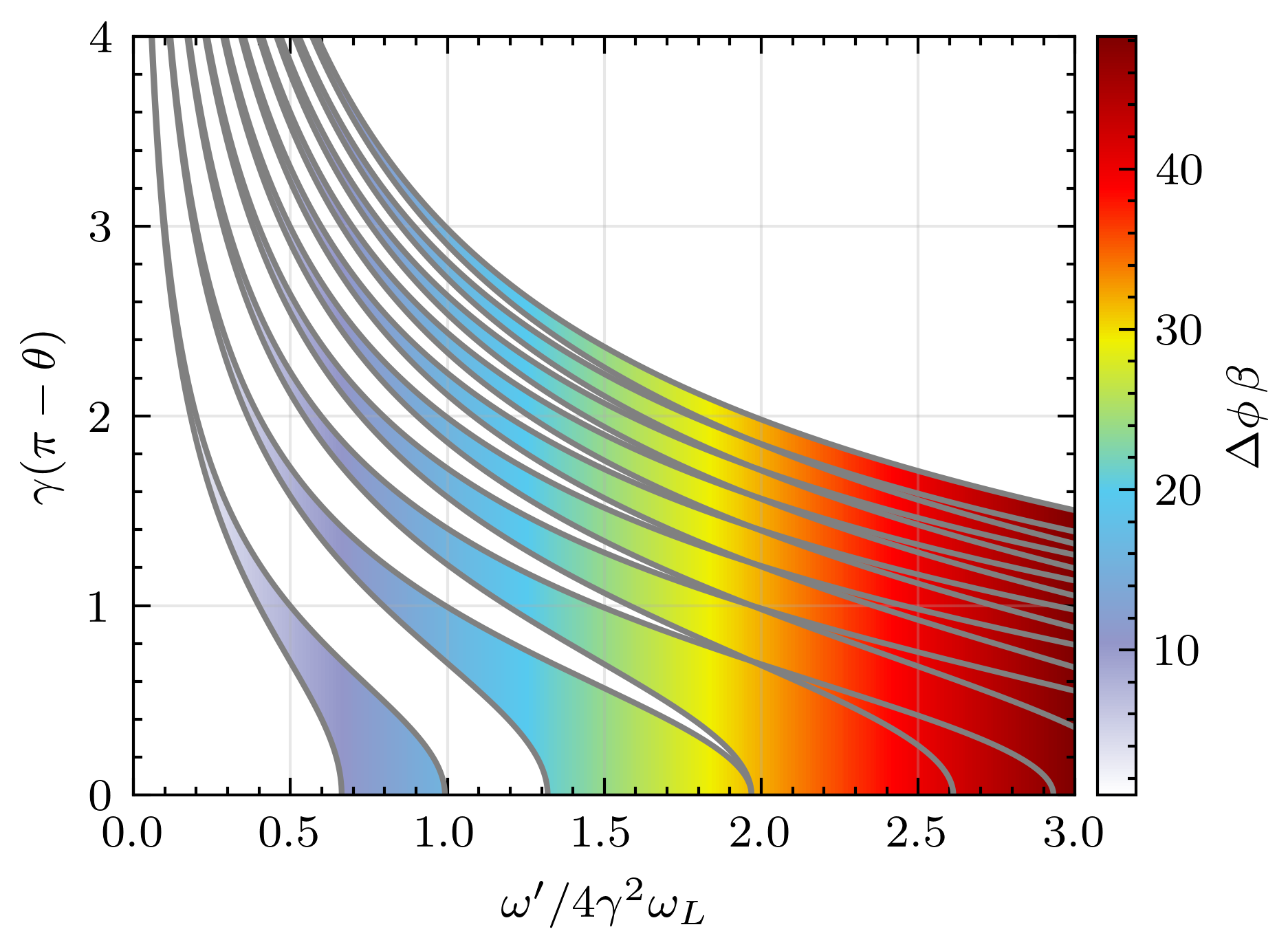}
    \caption{Asymptotic parameter $\Delta\phi\beta$ in the frequency--angle plane for
    $a_0=1$ and $\gamma=1000$. The gray lines indicate the boundaries of the first
    ten harmonic bands.}
    \label{fig:beta_map}
\end{figure}

\begin{figure*}[t]
    \centering
    \includegraphics[width=0.95\textwidth]{
    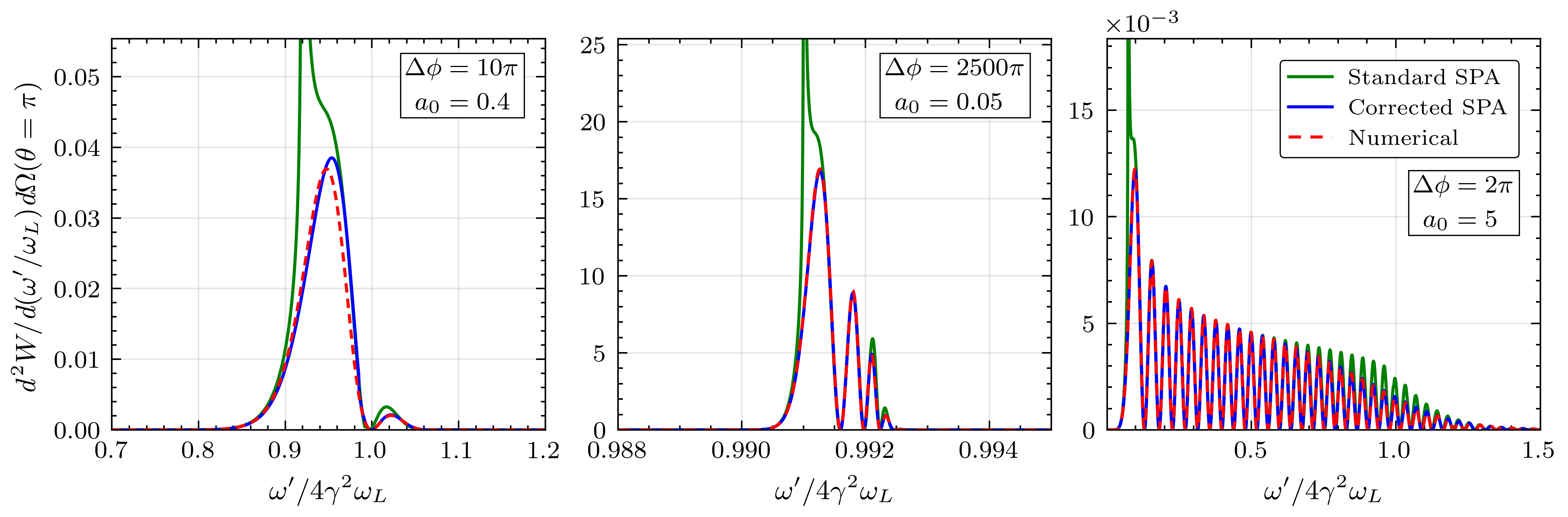}
    \caption{Backscattering spectra for circular polarization for different values of $a_0$ and $\Delta\phi$. The green, blue, and red dashed curves show the standard SPA, the corrected SPA, and the numerical calculation, respectively. The parameters are $\Delta\phi=10\pi$, $a_0=0.4$ (left panel), $\Delta\phi=2500\pi$, $a_0=0.05$ (center panel), and $\Delta\phi=2\pi$, $a_0=5$ (right panel), with $\gamma=1000$.}
    \label{fig:3_backscattering_spectra_circ}
\end{figure*}

The applicability of the asymptotic description is controlled by two related
but distinct conditions. The separation of the fast carrier oscillations from
the SVEA requires a multicycle pulse $\Delta\phi\gg1$, which is the usual SVEA condition. After this reduction, however, the
harmonic-resolved phase integrals \eqref{eq:C_l_n_main} are still governed by
the accumulated phase $\Delta\phi F_\ell(\phi)$. Thus a long pulse by itself is not sufficient for the saddle-point
approximation to be fully asymptotic if the nonlinear phase is too small. The
quantity $\Delta\phi\beta$ provides a useful diagnostic of the ponderomotive
phase accumulated over the pulse. When $\Delta\phi\beta \gg 1$, the phase varies rapidly between different emission phases, destructive interference suppresses the
nonstationary parts of the integral, and the saddle points determine the dominant contribution. When $\Delta\phi\beta \sim1$, the integral is not strongly localized near the saddle points, even if the pulse contains many cycles. For a head-on collision, and neglecting recoil corrections, $\beta$ scales as
\begin{equation}
    \beta \sim
    \frac{a_0^2\omega'}{4\gamma^2\omega_L}
    \sin^2\frac{\theta}{2}.
\end{equation}
This estimate shows that the accumulated nonlinear phase grows with the emitted photon frequency and with the laser intensity, and is largest in the
exact backscattering direction. Figure~\ref{fig:beta_map} shows the
distribution of $\Delta\phi\beta$ in the plane of the emitted photon frequency
and observation angle. The first ten harmonic bands are shown by the gray
curves. The parameter is largest in the high-frequency part of the spectrum,
where higher harmonics are located, and decreases when the observation
direction moves away from exact backscattering, both because of the angular
factor $\sin^2(\theta/2)$ and because the harmonic bands are shifted toward
lower photon frequencies. Therefore the SPA is expected to work best in the
regions where the accumulated nonlinear phase is large, and to deteriorate in
low-frequency angular regions where $\Delta\phi\beta$ is not large enough. This behavior is illustrated in
Fig.~\ref{fig:3_backscattering_spectra_circ} for circular polarization in the
backscattering direction. In the left panel, $a_0=0.4$ and $\Delta\phi=10\pi$. The pulse is multicycle, but the nonlinear broadening is weak and $\Delta\phi\beta$ is only of order unity. The spectrum is therefore
not yet fully in the saddle-point asymptotic regime, and visible deviations
from the numerical result remain. In the middle panel, the field is much
weaker $a_0=0.05$, but the pulse is very long $\Delta\phi=2500\pi$. Although the interaction is close to the linear regime,
the product $\Delta\phi a_0^2$ remains large enough for a sizable accumulated
nonlinear phase, and the finite-pulse subpeak structure is accurately
described by the SPA. The right panel shows the opposite limit. The pulse is
ultrashort, $\Delta\phi=2\pi$, but the intensity is large, $a_0=5$. This case
is outside the formal SVEA regime. However, in exact backscattering for
circular polarization the carrier-dependent phase vanishes, as discussed at
the end of Sec.~\ref{sec:master}, and the result is much less sensitive to the
SVEA reduction. The large value of $a_0$ produces strong ponderomotive
broadening and a large accumulated nonlinear phase. As a result, the corrected
SPA agrees well with the numerical calculation even for a very short pulse.

\begin{figure*}[t]
    \centering
    \includegraphics[width=0.95\textwidth]{
    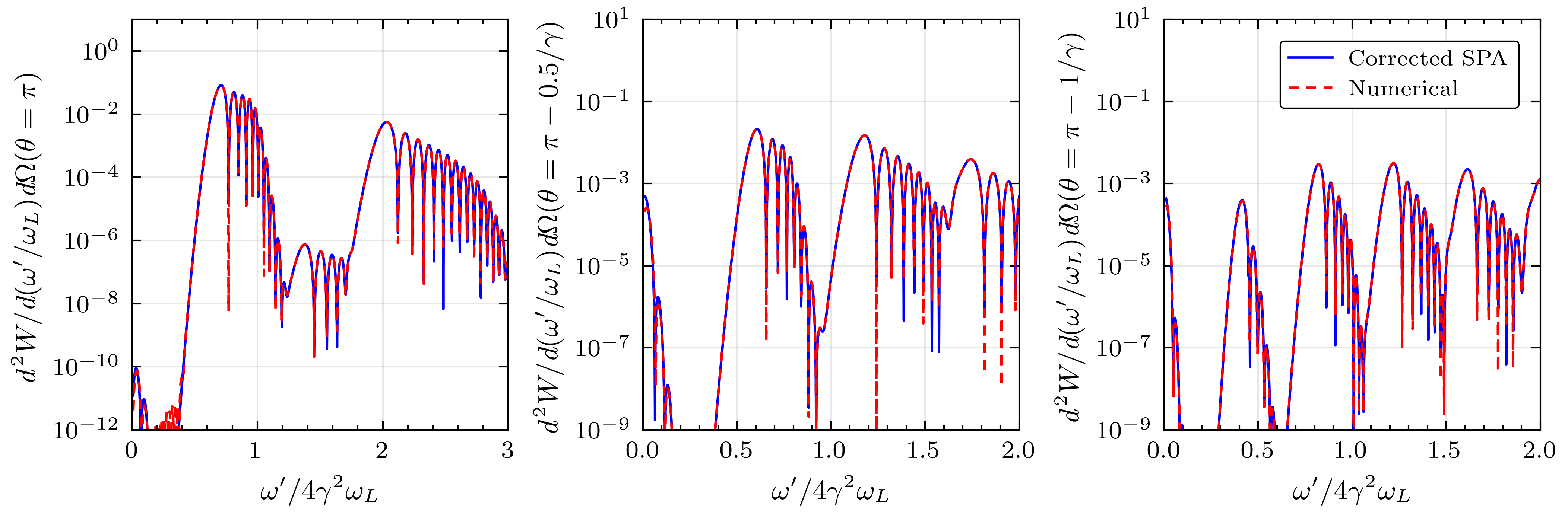}
    \caption{Spectra for linear polarization at different observation angles. The blue and red dashed curves show the corrected SPA and the numerical calculation, respectively. The parameters are $a_0=1$, $\Delta\phi=10\pi$, and $\gamma=1000$.}
    \label{fig:2_spectra_lin}
\end{figure*}

The angular dependence is shown in Fig.~\ref{fig:2_spectra_lin} for linear polarization. The spectra are plotted on a logarithmic scale for $\theta=\pi$, $\theta=\pi-0.5/\gamma$, and $\theta=\pi-1/\gamma$. In exact backscattering, the dominant contribution comes from odd harmonics, while the even harmonics are strongly suppressed. On the logarithmic scale one can still see a weak contribution of the second harmonic and a low-frequency contribution associated with the zeroth harmonic. The latter is not described accurately by the saddle-point formulae, because it lies at low photon frequencies where $\Delta\phi\beta$ is small. As the observation angle moves away from the axis, the low-frequency part becomes more visible and the agreement deteriorates first in this region. At the same time, for $a_0\sim1$ the corrected SPA still describes the main harmonic structure within an angular cone of order $1/\gamma$ around exact backscattering. This is the angular region where most of the radiation is emitted.
\begin{figure*}[t]
    \centering
    \includegraphics[width=0.95\textwidth]{
    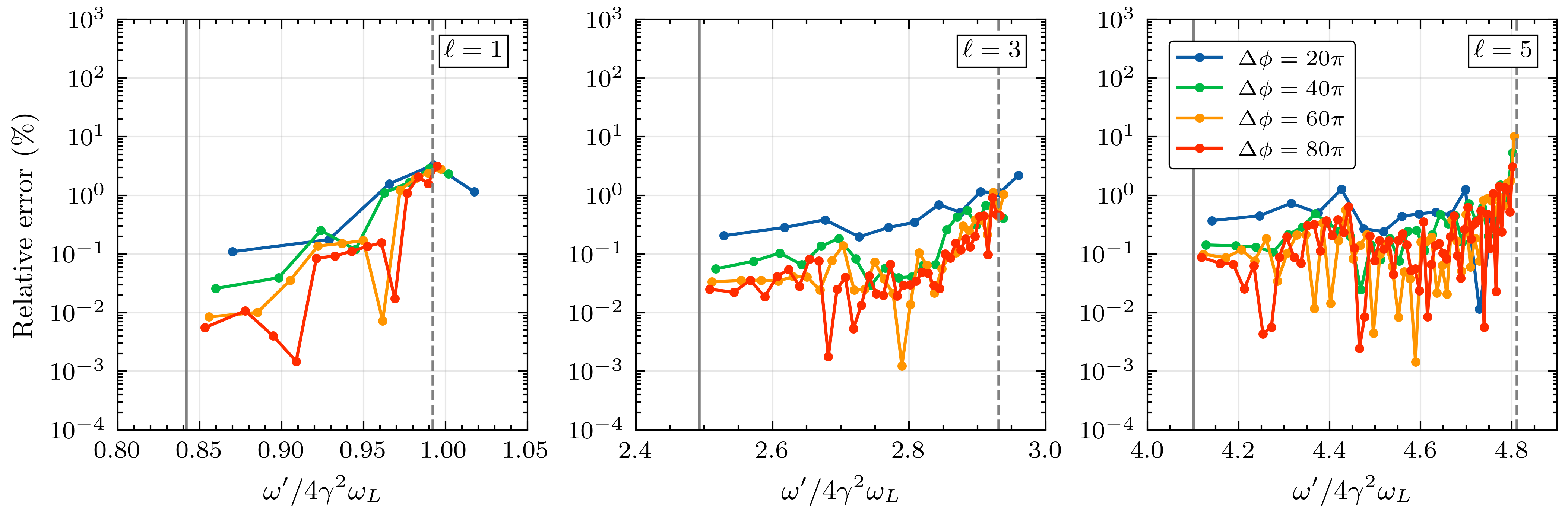}
    \caption{Relative error between the corrected SPA and numerical backscattering spectra for linear polarization, evaluated at the subpeak maxima for different pulse durations $\Delta\phi$. The three panels correspond to the odd harmonics $\ell=1$, $3$, and $5$. The parameters are $a_0=0.6$ and $\gamma=1000$.}
    \label{fig:accuracy}
\end{figure*}

We next test the accuracy of the corrected SPA formulae for
linear polarization in the exact backscattering direction. The results are shown in Fig.~\ref{fig:accuracy} for $a_0=0.6$, separately for the odd
harmonics $\ell=1$, $3$, and $5$. A pointwise comparison on a fixed frequency grid is not very informative in this case, because the spectrum contains zeros between neighboring subpeaks. Near such points even a small phase shift between the corrected SPA and numerical spectra leads to very large relative deviations. For this reason, the comparison is performed at the subpeak maxima in the region between the nonlinear and linear Compton edges. As the pulse duration increases, more subpeaks are resolved and the corrected SPA approaches
the numerical calculation better, which illustrates the asymptotic nature of the method. The accuracy, however, does not improve monotonically with the harmonic number. Although higher harmonics correspond to larger emitted frequencies and hence to a larger accumulated nonlinear phase, they are also weaker and depend more
sensitively on the generalized Bessel prefactors. As a result, the relative error for higher harmonics can remain comparable to that of the first harmonic. For all three harmonics, the largest deviations remain near the linear edge. This region is especially sensitive because the relevant saddle points lie in the weak-field tail of the pulse, where the result depends on
the details of the envelope falloff and on higher derivatives of the corrected phase. For longer pulses this sensitive region becomes narrower, and the corrected SPA describes most of each harmonic band with better accuracy.

\section{APPLICATION TO ELECTRON-BEAM SPECTRA}
\label{sec:beam}

The corrected SPA single-electron formulae can be used as building blocks for calculating incoherent radiation from an electron beam. In this case the total spectrum is obtained by summing the single-electron contributions over the sampled phase-space distribution of the beam. This provides a direct way to test whether the harmonic structure predicted for a single electron can survive after averaging over realistic beam parameters.

\begin{figure*}[t]
    \centering
    \includegraphics[width=0.95\textwidth]{
    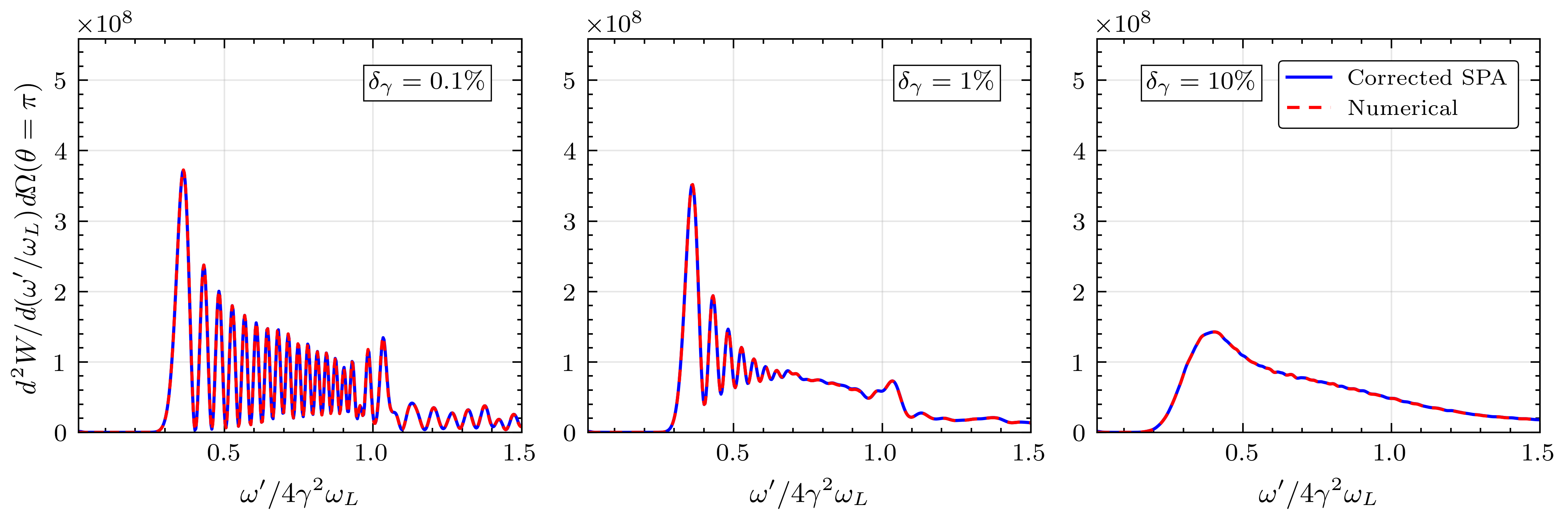}
    \caption{Incoherent backscattering spectra from an electron beam for a linearly polarized Gaussian pulse with $a_0=2$ and $\Delta\phi=10\pi$. The panels correspond to relative energy spreads $\delta_\gamma=0.1\%$, $1\%$, and $10\%$. The beam parameters are $Q=1\,\mathrm{nC}$, $\bar\gamma=100$, $\sigma_r=25\,\mu\mathrm{m}$, $\varepsilon=2.0\,\mathrm{mm\,mrad}$, and $N_{\mathrm{mp}}=10^4$. The blue and red dashed curves show the corrected SPA and the numerical calculation, respectively.}
    \label{fig:beam_spectra}
\end{figure*}

Figure~\ref{fig:beam_spectra} shows the incoherent backscattering spectrum for a linearly polarized Gaussian pulse with $a_0=2$ and $\Delta\phi=10\pi$. The beam charge is $Q=1\,\mathrm{nC}$, the mean Lorentz factor is $\bar\gamma=100$, the transverse rms size is $\sigma_r=25\,\mu\mathrm{m}$, and the transverse emittance is $\varepsilon=2.0\,\mathrm{mm\,mrad}$. The spectra are obtained using $N_{\mathrm{mp}}=10^4$ sampled macroparticles, which is sufficient to make the sampling fluctuations small on the scale shown in the figure. The three panels correspond to relative energy spreads $\delta_\gamma=0.1\%$, $1\%$, and $10\%$.

The first two values are motivated by the parameters of modern high-quality electron beams. Percent-level energy spreads have been demonstrated in laser-wakefield accelerators \cite{wiggins2010high}, while recent active energy-compression experiments have reached the sub-per-mille level \cite{winkler2025active}. For small energy spreads, $\delta_\gamma=0.1\%$ and $1\%$, the harmonic peaks and the finite-pulse subpeak structure remain visible even after summation over the beam. This demonstrates that the harmonic structure is not necessarily a single-electron effect, but can also be observable in the incoherent spectrum of a sufficiently high-quality electron beam.

As the energy spread increases, the spectra emitted by different electrons are shifted with respect to each other, and the subpeaks are progressively washed out. The transverse beam size $\sigma_r$ and emittance $\varepsilon$ can also affect the smearing of harmonic features through the transverse phase-space distribution, but here they are kept fixed in order to isolate the effect of the energy spread. For $\delta_\gamma=10\%$ the spectrum becomes almost smooth.

This behavior also clarifies the role of the present approximation. The standard SPA can resolve the finite-pulse substructure inside a harmonic band, but it fails near the harmonic edges. The LMA retains the local harmonic weights, but averages over the subpeak interference. The LCFA gives an even more local description and produces a smooth spectrum without harmonic substructure. In contrast, the corrected SPA keeps the coherent finite-pulse integrals and regularizes the edge regions. It can therefore resolve all visible subpeaks as long as they are not washed out by the electron-beam distribution. The agreement between the corrected SPA and numerical beam spectra demonstrates that the corrected SPA single-electron formulae can be used efficiently for calculating incoherent spectra after averaging over the electron-beam distribution. In particular, they provide an analytical tool for estimating when harmonic features should remain experimentally visible.

\section{CONCLUSIONS}
\label{sec:conclusion}

We have developed an analytical description of nonlinear Compton scattering in a finite plane-wave laser pulse with a smooth temporal envelope. Starting from the strong-field QED probability, the spectrum was reduced to phase integrals which contain the dependence on the pulse shape, polarization, and observation direction. In the multicycle regime, these integrals were organized into
harmonic-resolved phase integrals, which makes it possible to retain the coherent finite-pulse substructure of broadened harmonics.

The main result is a corrected asymptotic description of a harmonic band. In the central part of the band the spectrum is described by the ordinary two-saddle approximation, where the two saddles correspond to emission from the rising and falling parts of the pulse. At the nonlinear edge these saddles coalesce near the pulse maximum, and the resulting caustic is removed by a uniform Airy approximation. Near the linear edge the saddles move into the
weak-field tail, where the envelope factor must be included in the exponential. The corresponding envelope-corrected saddle-point analysis improves the description of the subpeaks near the linear edge.

We also showed how the locally monochromatic approximation is recovered within the same harmonic-resolved formulation. The LMA is obtained by averaging the finite-pulse interference encoded in the phase integrals, while keeping the local harmonic weights. In this way, the LMA preserves the averaged harmonic structure but removes the subpeak pattern produced by the coherent emission from different phases of the pulse. The LMA$^+$ partially restores finite-bandwidth effects and regularizes the local LMA caustic, but it remains
a local averaged approximation and does not recover the individual finite-pulse subpeaks.

The analytical formulae were compared with direct numerical calculations for circular and linear polarization, different pulse shapes, and different observation angles. We also applied the corrected SPA single-electron formulae to incoherent emission from an electron beam. The agreement with numerical calculations confirms that the method reproduces the harmonic edges, the dominant peaks, and the finite-pulse interference structure when the asymptotic parameter is sufficiently large. For sufficiently small energy spread, the harmonic substructure survives beam averaging and may therefore be observable in the incoherent spectrum of a high-quality electron beam.

The relation between the present approach and commonly used approximations is summarized in Table~\ref{tab:method_comparison}. The corrected SPA is most useful in the regime where the pulse contains many cycles and the harmonic substructure is experimentally or numerically resolved. Its applicability is limited by the assumptions of a smooth multicycle plane-wave pulse, a sufficiently large saddle-point parameter, and a simple two-saddle topology
near the nonlinear edge.

\begin{table*}[t]
\caption{Comparison of analytical and local approximations for finite-pulse nonlinear Compton spectra.}
\label{tab:method_comparison}
\begin{ruledtabular}
\begin{tabular}{p{0.18\textwidth}p{0.25\textwidth}p{0.25\textwidth}p{0.22\textwidth}}
Method & Harmonic structure & Edge behavior & Applicability \\
\hline

Corrected SPA&
Resolves harmonic bands and individual finite-pulse subpeaks &
Regularizes the nonlinear edge and improves the linear edge &
$\Delta\phi \beta \gg 1$; smooth envelope; two-saddle topology \\

Standard SPA
\cite{seipt2016analytical,kharin2016temporal,narozhnyi1996photon} &
Resolves subpeaks inside a harmonic band &
Diverges at the nonlinear edge and becomes inaccurate near the linear edge & $\Delta\phi \beta \gg 1$; isolated saddles\\

LMA
\cite{heinzl2020locally,larin2025extended} &
Resolves local harmonic bands, but averages over subpeaks &
Contains the local phase-density singularity at the nonlinear edge &
$\Delta\phi\gg1$; interference interval short compared with the envelope scale\\

LMA$^+$
\cite{larin2025extended} &
Restores a finite bandwidth of local harmonic lines &
Removes the fully differential LMA singularity for the chosen finite-width
profile &
LMA conditions; additionally
$\Delta\phi\gg2\pi a_0$ \\

LCFA
\cite{ritus1985quantum,di2018implementing,ilderton2019extended} &
Gives a smooth local spectrum without harmonic substructure &
Does not describe harmonic edges &
Short formation phase; typically $a_0\gg1$ and
$a_0^3/\chi\gg1$ \\

\end{tabular}
\end{ruledtabular}
\end{table*}

\begin{acknowledgments}
E.G.G., O.K., and S.W. acknowledge support from the Czech Science Foundation (GAČR), project No.~24-14395L. Th.~B. was supported by the ELI Beamlines internship program (2025). S.G.R. and A.M.F. were supported by the National Center for Physics and Mathematics under research program No.~6. M.P.M. was supported by Project No.~FSWU-2026-0009 of the Ministry of Science and Higher Education of the Russian Federation.
\end{acknowledgments}

\appendix

\section{Thomson limit}
\label{app:thomson}
The classical Thomson limit corresponds to negligible recoil of the electron. It is convenient to parameterize this limit by the invariant ratio
\begin{equation}
    \rho
    =
    \frac{k\cdot p'}{k\cdot p}
    =
    1-
    \frac{k\cdot k'}{k\cdot p},
    \label{eq:rho_thomson}
\end{equation}
where $0\leq\rho\leq1$. The quantity $\rho$ measures the fraction of the
initial value of $k\cdot p$ retained by the final electron, while
$1-\rho$ is the corresponding fraction carried by the emitted photon. The low-recoil limit is therefore $1-\rho\ll1$. In terms of the variable
$u$ used in the probability formula \eqref{eq:dW_main}, this gives
$u=(1-\rho)/\rho\ll1$, and the momentum transfer variable reduces to
$s=(p\cdot k')/(k\cdot p)$.

In this limit the recoil-dependent coefficients \eqref{eq:alpha}-\eqref{eq:beta} entering the phase can be
expanded in powers of $1-\rho$. To leading order one obtains
\begin{align}
    \alpha_{1,2}
    &=
    \frac{ma_0}{k\cdot p}
    \left[
        \frac{k\cdot k'}{k\cdot p}
        \left(\varepsilon_{1,2}\cdot p\right)
        -
        \varepsilon_{1,2}\cdot k'
    \right],
    \label{eq:alpha_thomson}
    \\
    \beta
    &=
    \frac{m^2a_0^2}{4}
    \frac{k\cdot k'}{(k\cdot p)^2}.
    \label{eq:beta_thomson}
\end{align}

Applying the low-recoil expansion to the spectral-angular energy distribution
\eqref{eq:dE_main} gives
\begin{multline}
  \frac{\dd^{2} \mathcal{E}}{\dd\omega'\dd\Omega}
  =
  \frac{\alpha m^2\omega'^2}
       {8\pi^{2}(k\cdot p)^2}
  \Bigg[
      -2|A_0|^2
      +
      \frac{a_0^2}{2}
      \bigg(
          |A_+|^2
          +
          |A_-|^2
          \\[1mm]
          +
          2\cos 2\xi\,\operatorname{Re}\left[A_+A_-^*\right]
          -
          2\operatorname{Re}\left[A_0A_2^*\right]
      \bigg)
  \Bigg].
  \label{eq:dE_thomson}
\end{multline}

\section{Uniform approximation at the nonlinear edge}
\label{app:uniform}

We use the canonical-integral method for two coalescing saddle points
\cite{Felsen_Marcuvitz_RadiationScattering_1994,milovsevic2025application}. The original phase is mapped to the cubic canonical phase by an implicit
change of variables:
\begin{equation}
    iF_\ell(\phi)
    =
    \tau(z),
    \qquad
    \tau(z)
    =
    \eta+\sigma z-\frac{z^3}{3}.
    \label{eq:tau_def_app}
\end{equation}
The saddle points of the canonical phase are $z_{1,2}=\pm\sqrt{\sigma}$. Matching the phase values at the original saddle points $\phi_{1,2}$ gives
\begin{align}
    \eta
    &=
    \frac{i}{2}
    \left(
        F_\ell(\phi_1)+F_\ell(\phi_2)
    \right), \\
    \sigma^{3/2} 
    &=
    \frac{3i}{4}
    \left(
        F_\ell(\phi_1)-F_\ell(\phi_2)
    \right).
    \label{eq:eta_sigma_app}
\end{align}
The branch of $\sigma$ is chosen so that the steepest-descent contour and the quantities defined below vary continuously through the coalescence point.

\begin{figure*}[t]
    \centering
    \includegraphics[width=0.95\textwidth]{
    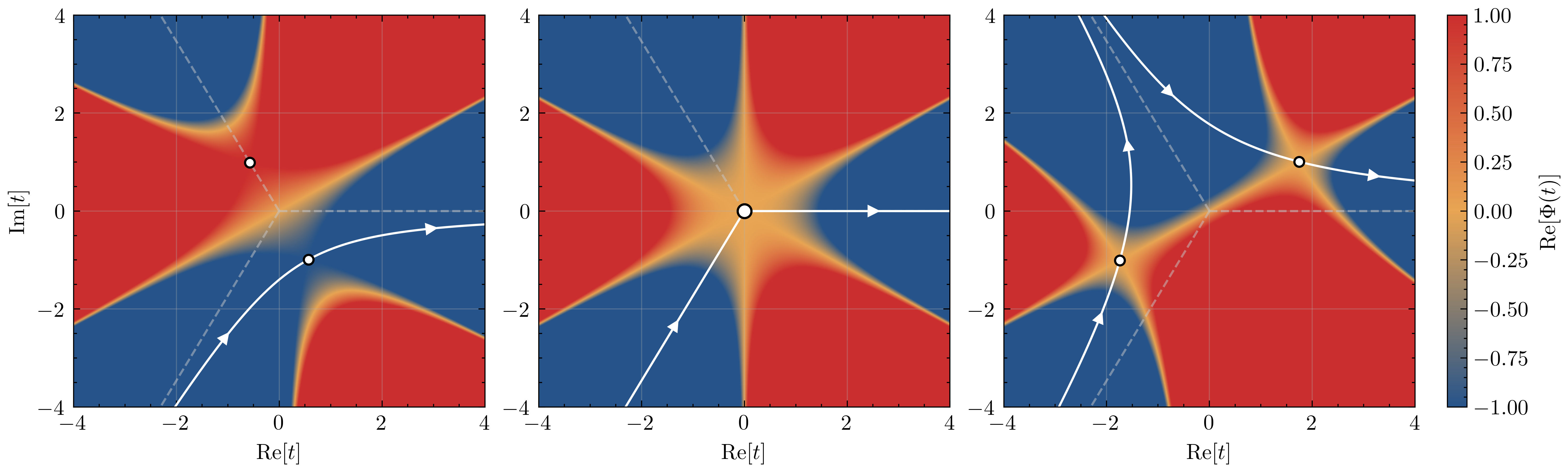}
    \caption{Steepest-descent contours in the canonical Airy plane. The panels correspond to $s<s_\ell$, $s=s_\ell$, and $s>s_\ell$. The background shows
    $\operatorname{Re}[\Phi(t)]$. Open circles mark the saddle points and white curves show the relevant contours.}
    \label{fig:airy_contours}
\end{figure*}

After the change of variables, we introduce the transformed prefactor
\begin{equation}
Q(z)=\mathcal{J}_{r}\!\left(\phi(z)\right)g^n\!\left(\phi(z)\right)\frac{d\phi}{dz}.
\end{equation} 
In the leading uniform approximation, this prefactor is replaced by the
linear function $Q(z)\approx K_{+}+K_{-} z$ that reproduces its values at the two stationary points, where
\begin{equation}
    K_{+} = \frac{Q(z_1)+Q(z_2)}{2},
    \qquad
    K_{-} = \frac{Q(z_1)-Q(z_2)}{2\sqrt{\sigma}}.
\end{equation}
This retains the contributions associated with both saddles and remains
regular when they coalesce. After the rescaling $t=\Delta\phi^{1/3}z$, the exponential is reduced to the canonical Airy phase $\Phi(t)=\zeta t - t^3/3$ with $\zeta=\sigma\Delta\phi^{2/3}$. Keeping this leading uniform term gives
\begin{equation}
    \mathcal{C}_{n, r}^{(\ell)}
    \approx
    \frac{e^{\eta\Delta\phi}}{\Delta\phi^{1/3}}
    \left(K_{+}C(\zeta)
    +
    \frac{K_{-}C'(\zeta)}{\Delta\phi^{1/3}}
    \right),
    \label{eq:DN_uniform_general_app}
\end{equation}
where the canonical Airy-type integral is defined as
\begin{equation}
    C(\zeta)
    =
    \int_P
    \exp\left(\zeta t-\frac{t^3}{3}\right)\dd t.
    \label{eq:C_def_app}
\end{equation}
Here $P$ is one of the admissible Airy contours connecting two decay sectors of the cubic exponential \cite{abramowitz1964handbook} and $C'(\zeta)=dC/d\zeta$. The relevant contour is obtained by continuously
deforming the original real integration path under the mapping
\eqref{eq:tau_def_app}. This contour prescription also fixes the branch of the cubic root entering $\sigma$. The evolution of the steepest-descent contour in the canonical plane is shown
in Fig.~\ref{fig:airy_contours}. The three panels correspond to the regions
before, at, and after the coalescence of the saddle points. The figure
illustrates how the relevant contour is selected by continuous deformation of
the original real integration path.

For an even envelope $g(-\phi)=g(\phi)$, with a single nondegenerate maximum at $\phi=0$, the two relevant saddle points form the symmetric pair $\phi_{1,2}=\pm\phi_0$, where $\phi_0$ may be either real or complex depending on the spectral region. The phase is antisymmetric at the two saddles $F_\ell(-\phi_0)=-F_\ell(\phi_0)$, while the envelope and the remaining factors in the prefactor have equal values. Consequently $\eta=0$, and the canonical parameter can be determined from either member of the saddle pair. We therefore define
\begin{equation}
    \zeta(\phi)
    =
    \left[
        \frac{3i}{2}\Delta\phi\,F_\ell(\phi)
    \right]^{2/3}.
    \label{eq:zeta}
\end{equation}
We denote by $|\zeta_0|=|\zeta(\phi_0)|$ the magnitude associated with the symmetric saddle pair.  The Jacobians at the two canonical stationary points are obtained by differentiating the phase mapping twice and evaluating it at the corresponding saddles,
\begin{equation}
    \left.
    \frac{d\phi}{dz}
    \right|_{z=\pm\sqrt{\sigma}}
    =
    \sqrt{
        \frac{\mp 2\sigma^{1/2}}
             {iF_\ell''(\phi_{1,2})}
    } .
    \label{eq:h12_app}
\end{equation}
The signs of the square roots are fixed by the local orientation of the
corresponding steepest-descent contours. With this choice, the symmetry of the saddle pair gives equal values of the transformed prefactor at the two stationary points. Hence $K_{-}=0$, and the term proportional to $C'(\zeta)$ in Eq.~\eqref{eq:DN_uniform_general_app} vanishes. It is then convenient to express the Airy prefactor through the mapping $\zeta(\phi)$. Differentiating Eq.~\eqref{eq:zeta} twice and using $F_\ell'(\phi_0)=\zeta'(\phi_0)=0$ gives the local matching condition
quoted in Eq.~\eqref{eq:zeta_second_derivative}. With the contour prescription
described above, this leads to the uniform approximation \eqref{eq:DN_uniform_airy}.

\section{Envelope-corrected linear-edge expansion}
\label{app:linear_edge}

\begin{figure}[t]
    \centering
    \includegraphics[width=0.95\columnwidth]{
    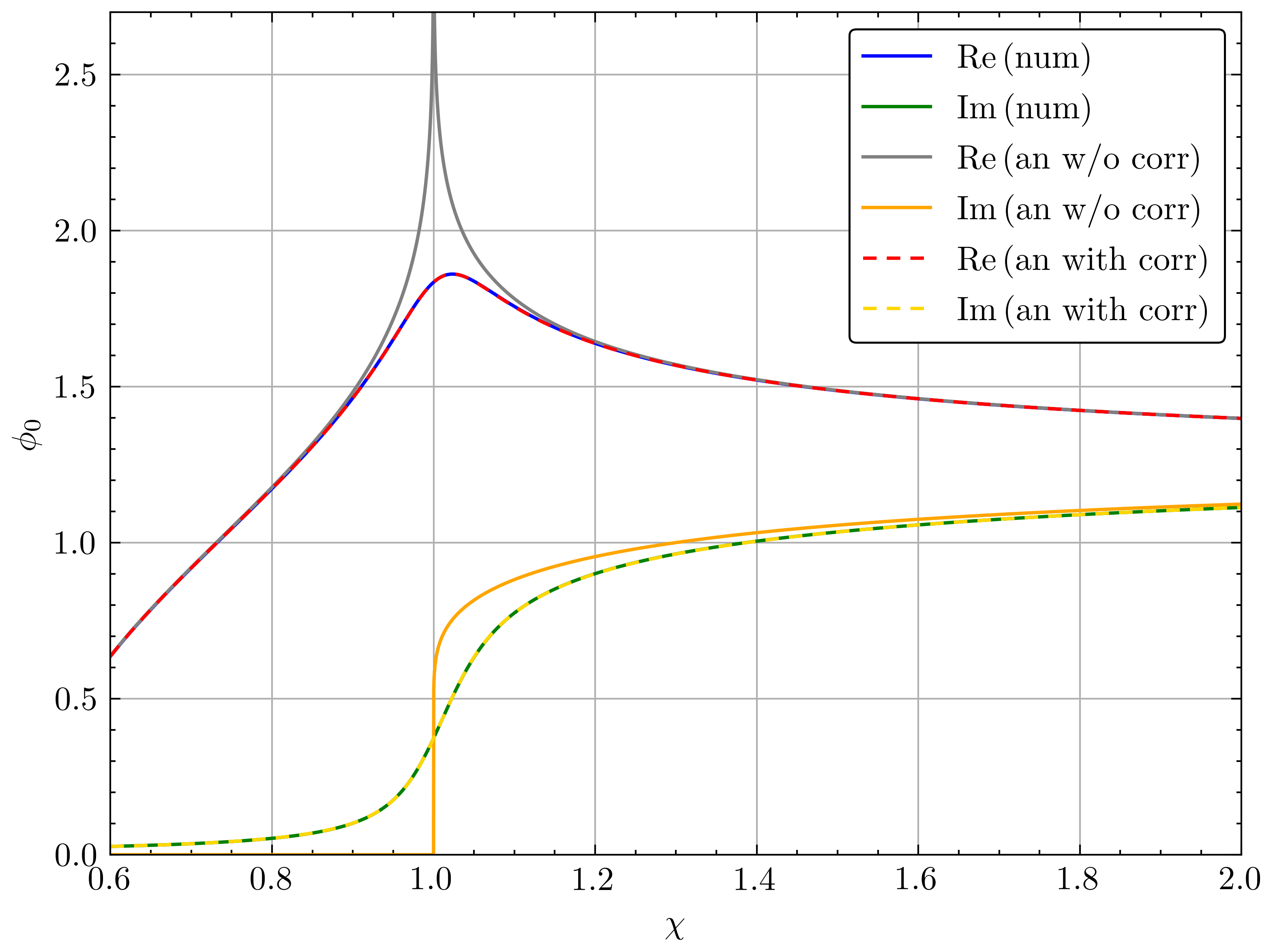}
    \caption{Saddle point $\phi_0$ in the vicinity of $s \approx \ell$ for a Gaussian pulse. The solid blue and green curves show the real and imaginary parts obtained numerically with full inclusion of the envelope. The gray and orange curves correspond to the analytical solution neglecting envelope effects. The dashed red and yellow curves represent the approximate analytical solution including the envelope correction.}
    \label{fig:linear_edge_saddle}
\end{figure}

In this Appendix we describe how the corrected saddle points used near the linear edge are constructed. The starting point is the envelope-corrected phase $q_{\ell,n}$ and the corresponding saddle-point equation $q'_{\ell,n}(\phi_0)=0$, introduced in Eq.~\eqref{eq:qN_corr}. The additional logarithmic term in $q_{\ell,n}$ is formally suppressed by $1/\Delta\phi$, but it becomes essential near $s=\ell$, where the uncorrected saddle moves into the weak-field tail of the pulse. At the linear edge $s=\ell$, the leading envelope-controlled saddle is found from
\begin{equation}
    \bar\phi_0=H^{-1}\!\left(\frac{\Delta\phi b\ell}{in}\right),
    \qquad
    H(\phi)=\frac{g'(\phi)}{g^3(\phi)}.
    \label{eq:phi0_edge_app}
\end{equation}
For $s$ close to the edge we use the interpolating zeroth-order approximation
\begin{equation}
    \tilde\phi_0=g^{-1}
    \left(
        \left[
            \frac{\ell-s}{\beta}
            +
            g^2(\bar\phi_0)e^{-(\ell-s)^2}
        \right]^{1/2}
    \right).
    \label{eq:phi0_fit_app}
\end{equation}
Although Eq.~\eqref{eq:phi0_fit_app} is an interpolation away from the edge, it gives the exact envelope-controlled saddle at $s=\ell$, where $\tilde\phi_0=\bar\phi_0$. In the interior of a harmonic band, it approaches the ordinary saddle position when the envelope correction becomes negligible. The final saddle used in the linear-edge approximation is obtained by one Newton step applied to the corrected phase,
\begin{equation}
    \phi_0=\tilde\phi_0-
    \frac{q'_{\ell,n}(\tilde\phi_0)}
         {q_{\ell,n}^{(2)}(\tilde\phi_0)} .
    \label{eq:phi0_corr_app}
\end{equation}
The correction is small in the overlap region, because the interpolating position already lies close to the numerical solution of the corrected saddle equation. This is illustrated in Fig.~\ref{fig:linear_edge_saddle}.

The inverse functions entering Eq.~\eqref{eq:phi0_edge_app} and
Eq.~\eqref{eq:phi0_fit_app} depend on the pulse shape. For the envelopes used
in the numerical examples they are listed in Table~\ref{tab:pulse_shapes}.
\begin{table}[t]
\centering
\caption{Envelope functions used in the linear-edge approximation.}
\label{tab:pulse_shapes}
\begin{tabular}{lccc}
\hline\hline
Envelope & \(g(\phi)\) & \(g^{-1}(y)\) & \(H^{-1}(y)\) \\
\hline
Gaussian
&
\(e^{-\phi^2/2}\)
&
\(\sqrt{2\ln(1/y)}\)
&
\(\left(\dfrac{1}{2}W(2y^2)\right)^{1/2}\)
\\[2mm]
Hyperbolic secant
&
\(\operatorname{sech}\phi\)
&
\(\operatorname{arccosh}(1/y)\)
&
\(-\dfrac{1}{2}\operatorname{arcsinh}(2y)\)
\\
\hline\hline
\end{tabular}
\end{table}
Here $W$ is the Lambert function. The inverse functions are understood with the branches which continuously connect the corrected saddle to the relevant steepest-descent contour. For the hyperbolic-secant envelope the corrected saddle-point equation \eqref{eq:qN_corr} can be solved exactly for arbitrary $s$. Introducing the variable $v=\tanh\phi$ reduces this equation to an algebraic
one, and the corresponding saddle can be written as
\begin{equation}
    \phi_{1,2}
    =
    \operatorname{artanh}\left(\frac{-n \pm \sqrt{n^2-4\Delta\phi^2\beta(\beta+s-\ell)}}{2i\Delta\phi\beta}\right).
    \label{eq:sech_exact_saddle_app}
\end{equation}
This exact solution is useful as a check of the interpolating construction above.

We next derive the higher-order correction appearing in
Eq.~\eqref{eq:DN_linear_main}. In a neighborhood of the corrected saddle with $t=\phi-\phi_0$, we write
\begin{align}
    q_{\ell,n}(\phi)
    &\approx
    q_0
    +
    \frac{q_2}{2}t^2
    +
    \frac{q_3}{6}t^3
    +
    \frac{q_4}{24}t^4,
    \label{eq:q_expansion_linear_app}
    \\
    \mathcal{J}_{r}(\phi)
    &\approx
    \mathcal{J}_{r,0}
    +
    \mathcal{J}'_{r,0}t
    +
    \frac{\mathcal{J}''_{r,0}}{2}
    t^2,
    \label{eq:J_expansion_linear_app}
\end{align}
where $q_i=q_{\ell,n}^{(i)}(\phi_0)$, $\mathcal{J}_{r,0}^{(i)}=\mathcal{J}_{r}^{(i)}(\phi_0)$. The quadratic part of the phase determines the Gaussian saddle contribution. The cubic and quartic phase terms, together with the first and second derivatives of the prefactor, generate the first correction in
$1/\Delta\phi$.

The characteristic width of the saddle region is $\delta\phi \sim \left(\Delta\phi\,|q_2|\right)^{-1/2}$. The perturbative treatment of the higher phase derivatives requires
\begin{align}
    \frac{|q_3|}{\Delta\phi^{1/2}|q_2|^{3/2}}
    &\ll 1,
    &
    \frac{|q_4|}
         {\Delta\phi|q_2|^2}
    &\ll 1.
    \label{eq:linear_validity_app}
\end{align}
These conditions become increasingly accurate for long pulses and away from regions where the quadratic saddle approximation itself becomes degenerate.

Evaluating the Gaussian moments with the expansions
\eqref{eq:q_expansion_linear_app} and \eqref{eq:J_expansion_linear_app}
gives the correction factor $\Sigma$ quoted in
Eq.~\eqref{eq:linear_defs_main}. The terms containing
$\mathcal{J}'_{r,0}$ and $\mathcal{J}''_{r,0}$ come from the variation of the
generalized Bessel factor across the saddle region, while the terms containing
$q_3$ and $q_4$ are the post-Gaussian corrections generated by the higher
derivatives of the corrected phase.

The square root in the single-saddle contribution is chosen according to the
orientation of the steepest-descent contour through $\phi_0$. For an even
envelope, the second contributing saddle is $-\phi_0^*$. The corrected phase
and the prefactor then satisfy
\begin{equation}
    q_{\ell,n}^{(i)}(-\phi_0^*)
    =
    \left[
        q_{\ell,n}^{(i)}(\phi_0)
    \right]^*,
    \quad
    \mathcal{J}_{r}^{(i)}(-\phi_0^*)
    =
    \left[
        \mathcal{J}_{r}^{(i)}(\phi_0)
    \right]^*.
\end{equation}
Consequently, the two saddle contributions are complex conjugates and combine
into the real oscillatory expression given in Eq.~\eqref{eq:DN_linear_main}.
The exponential factor in that expression describes the suppression associated
with the weak-field tail of the pulse, while
$(\Delta\phi|q_2|)^{-1/2}$ determines the local formation scale. The cosine
comes from the interference of the two conjugate saddle contributions.

Away from the linear edge, the logarithmic term in Eq.~\eqref{eq:qN_def}
becomes a small correction over the saddle region. The corrected saddle then
approaches the ordinary one, and Eq.~\eqref{eq:DN_linear_main} reduces to the
standard two-saddle approximation.


%

\end{document}